\magnification=\magstep1
\voffset=-0.5 true in
\vsize=8truein
\baselineskip=24pt
\pageno=1
\pretolerance=10000

\def\etal{{\it et al.}\hskip 1.5pt}
\def\cf{{\it cf.}\hskip 1.5pt}
\def\refset{\parindent=0pt\hangafter=1\hangindent=1em}

\def\lb{\left(}
\def\rb{\right)}
\def\x{{\bf x}}
\def\rb{\right)}
\def\lb{\left(}

\def\u{{\bf u}}

\def\p{{\bf p}}

%%%%%%%%%%% MACRO BEGINS %%%%%%%%%%%%%%%%%%

\newcount\equationno      \equationno=0
\newtoks\chapterno \xdef\chapterno{}
\def\eqn{\eqno\eqname}
\def\eqname#1{\global \advance \equationno by 1 \relax
\xdef#1{{\noexpand{\rm}(\chapterno\number\equationno)}}#1}

% less than or order of \la

% greater than or order of \ga
\def\ga{\mathrel{\mathchoice {\vcenter{\offinterlineskip\halign{\hfil
$\displaystyle##$\hfil\cr>\cr\sim\cr}}}
{\vcenter{\offinterlineskip\halign{\hfil$\textstyle##$\hfil\cr>\cr\sim\cr}}}
{\vcenter{\offinterlineskip\halign{\hfil$\scriptstyle##$\hfil\cr>\cr\sim\cr}}}
{\vcenter{\offinterlineskip\halign{\hfil$\scriptscriptstyle##$\hfil\cr>\cr\sim\cr}}}}}
%
% +--------------------------------------------------------------------+
% |								       |
% |			      TABLES.TEX			       |
% |								       |
% |			Ray F. Cowan  15-Feb-85			       |
% |								       |
% |			  Princeton University			       |
% |								       |
% |			Last Revision: 6-May-85			       |
% |								       |
% |   Macros I find handy for making tables.  See TABLEDOC TEX for     |
% |   a	longer description.  The token-counting	macros are straight    |
% |   from the TeXbook's "Dirty	Tricks"	appendix.		       |
% |								       |
% +--------------------------------------------------------------------+
%
\newbox\hdbox%
\newcount\hdrows%
\newcount\multispancount%
\newcount\ncase%
\newcount\ncols% This is the number of primary text columns in the table.
\newcount\nrows%
\newcount\nspan%
\newcount\ntemp%
\newdimen\hdsize%
\newdimen\newhdsize%
\newdimen\parasize%
\newdimen\thicksize%
\newdimen\thinsize%
\newdimen\tablewidth%
\newif\ifcentertables%
\newif\ifendsize%
\newif\iffirstrow%
\newif\iftableinfo%
\newtoks\dbt%
\newtoks\hdtks%
\newtoks\savetks%
\newtoks\tableLETtokens%
\newtoks\tabletokens%
\newtoks\widthspec%
%
%  Book-keeping	stuff--see how often these macros are called.
%
% 09/17/86 The three lines below were changed as the next three	by H.M.
%\immediate\write15{%
%CP SMSG GJMSINK TEXTABLE -----> TABLE MACROS LOADED, JOB = \jobname%
%}%
\immediate\write15{%
-----> TABLE MACROS LOADED%
}%
%
%  Turn	on table diagnostics.
%
\tableinfotrue%
\catcode`\@=11%	 Allows	use of "@" in macro names, like	PLAIN.TEX does.
%  Debugging aid.  Writes #1 on the
%				     user's terminal and in the	log file.
\def\tstrut{\vrule height16pt depth6pt width0pt}%
\def\|{|}%  Make it easy to get	|'s of type other after	they are later
%	    made active.
\def\tablerule{\noalign{\hrule height\thinsize depth0pt}}%
\thicksize=1.5pt%  Default thickness for fat rules.  The user should feel
%		   free	to change this to his preference.
\thinsize=0.6pt%   Default thickness for thin rules.
\def\thickrule{\noalign{\hrule height\thicksize	depth0pt}}%
\def\ctr#1{\hfil\ #1\hfil}%
%
%
%  Here	are things for controlling the width of	the finished table.
%
\tablewidth=-\maxdimen%
\def\tabskipglue{0pt plus 1fil minus 1fil}%
%
%  Stuff for centering or not.
%
\centertablestrue%
%
%
%
%  \vctr vertically centers its	argument in the	row.
%
\parasize=4in%
\gdef\ARGS{########}%  Produces	the correct number of #'s in the preamble
%		       by the time everything is expanded and \halign sees
%		       it.
\gdef\headerARGS{####}%	 Same as \ARGS,	but used in \header macros.
\def\@mpersand{&}%  Allows us to get alignment tab characters later
%		    when we have made the character "&"	an active macro.
{\catcode`\|=13%  Make |'s locally active.
\gdef\letbarzero{\let|0}%  Globally define a macro that	allows us to
%			   keep	active |'s from	being expanded in edef's.
\gdef\letbartab{\def|{&&}}%  This \def will cause active |'s read by
%			     \ruledtable to be converted into double
%			     alignment tabs.
}%  End	of locally active |'s.
{\def\ampskip{&\omit\hfil&}%  This local macro skips a vertical	rule.
\catcode`\&=13%	 Now make &'s into active macros.
\let&0%	 This allows us	to expand \ampskip in the next \xdef without
%	 attempting to expand the & and	getting	an "undefined control
%	 sequence" error.
\xdef\letampskip{\def&{\ampskip}}%  This will cause active &'s read by
%				    \ruledtable	to be converted	into
%				    double tabs	and an \omit'ted \vrule.
}%  End	of locally active &'s.
\def\begintable{%  Here	we make	|'s and	&'s active characters so we can
%		   interpret them as macros.  Note that	this action is
%		   true	only until we encounter	the matching \endgroup
%		   token later at the end of the \ruledtable macro.
   \begingroup%
   \catcode`\|=13\letbartab%
   \catcode`\&=13\letampskip%
   \def\multispan##1{%	We must	redefine \multispan to count the number
%			of primary columns, not	physical columns.
      \omit \mscount##1%
      \multiply\mscount\tw@\advance\mscount\m@ne%
      \loop\ifnum\mscount>\@ne \sp@n\repeat%
   }%  End of \multispan macro.
   \def\|{%
      &\omit\widevline&%
   }%
   \ruledtable%	 Now we	call \ruledtable to do the real	work.
}%  End	of \begintable macro.
\long\def\ruledtable#1\endtable{%
%
%  This	macro reads in the user's data entries
%  and converts	them into a ruled table.
%
%  Important note:  Many macros	and parameters are re-defined here, and
%  these must be kept local to the table macros	to avoid conflict with
%  their use outside of	tables.	 This is done by the \begingroup token
%  macro \begintable and the \endgroup token at	the end	of
%  this	macro.
%
   \offinterlineskip%  Needed to make rules touch each other.
   \tabskip 0pt%  Needed for same reason as \offinterlineskip.
   \def\widevline{\vrule width\thicksize}%  Make outer \vrule's	wider.
   \def\endrow{\@mpersand\omit\hfil\crnorm\@mpersand}%
   \def\crthick{\@mpersand\crnorm\thickrule\@mpersand}%
   \def\crnorule{\@mpersand\crnorm\@mpersand}%
   \let\nr=\crnorule%  A shorter abbreviation.
   \def\endtable{\@mpersand\crnorm\thickrule}%
   \let\crnorm=\cr%  Allows us to use \cr for our own purposes.
%
%  Cause user-typed \cr's to follow a row with a \tablerule.
%
   \edef\cr{\@mpersand\crnorm\tablerule\@mpersand}%
   \the\tableLETtokens%	 Get the user's	extra \let's, if any.
%
%  Put the data	entries	into a token register so we can	scan through them
%  and see what	the user is asking us to do.
%
   \tabletokens={&#1}%	We add an extra	alignment tab to the beginning
%			of the first row to allow for the first	\vrule.
%
%  Now count how many rows are in the table and	return the result in
%  count register \nrows; do the same for columns, and return that
%  in register \ncols.
%
   \countROWS\tabletokens\into\nrows%
   \countCOLS\tabletokens\into\ncols%
%
%  Now do a little arithmetic to convert the number of primary columns
%  into	the number of physical columns that the	alignment preamble must
%  prepare for;	 similarly for rows.
%
   \advance\ncols by -1%
   \divide\ncols by 2%
   \advance\nrows by 1%
%
%  Tell	the user how many rows and columns we found in his data, if he
%  wants to know.
%
   \iftableinfo	%
      \immediate\write16{[Nrows=\the\nrows, Ncols=\the\ncols]}%
   \fi%
%
%  Now we actually go ahead and	produce	the table.
%
   \ifcentertables
      \line{%  The final table comes out as an \hbox of	width the \hsize.
      \hss%  The final table will be centered left-to-right.
   \else %
      \hbox{%
   \fi
      \vbox{%
	 \makePREAMBLE{\the\ncols}%  Generate the preamble.
	 \edef\next{\preamble}%	 This line and the next	line force the
	 \let\preamble=\next%	 expansion of all \ARGS	tokens into the
%				 appropriate number of #'s.
	 \makeTABLE{\preamble}{\tabletokens}%  Go do the \halign here.
      }%  End of \vbox.
      \ifcentertables \hss}\else }\fi%	Finish the centering effect.
%					It is important	that no	spaces
%					follow the two `}' here.
%  }%  End of \line.
   \endgroup%  Return all local	macros and parameters to their outside
%	       values.
   \tablewidth=-\maxdimen%  Reset \tablewidth to normal.
}%  End	of macro \ruledtable.
\def\makeTABLE#1#2{%  Does an \halign for the \ruledtable macro.
   {%  Start of	local parameter	values.
   \let\ifmath0%     These macros would	cause trouble if they were to be
   \let\header0%     expanded in the following \xdef; we \let them be
   \let\multispan0%  equal to a	digit, because digits can't be expanded.
%
%  Set up the width specification here.
%
   \ifdim\tablewidth>-\maxdimen	%
 \widthspec=\expandafter{\expandafter t\expandafter o%
 \the\tablewidth}%
   \else %
      \widthspec={}%
   \fi %
%\out{Widthspec=[\the\widthspec]}%
%\out{Preamble=[\preamble]}%
   \xdef\next{%	 We must force the preamble to be expanded BEFORE the
      \halign\the\widthspec{%
%	 \halign is done;  this	\edef\next{...}\next construction
%		 does the trick.
      #1%  This	is the preamble	text.
      \noalign{\hrule height\thicksize depth0pt}%  Makes the top \hrule.
      \the#2\endtable%	This is	the main body.
%
%     \noalign{\hrule height0.7pt depth0pt}%  Makes the	last \hrule.
      }%  End of \halign.
   }%  End of \next.
   }%  End of local values.
   \next%  This	\next must be outside of the local values, because now
%	   we want those troublesome macros in the \let's above	to have
%	   their normal	actions.
}%  End	of macro \makeTABLE.
\def\makePREAMBLE#1{%  This macro generates the	necessary preamble for a
%		       ruled table with	#1 primary columns.
%		       (Primary	columns	means the number of columns NOT
%			counting those used for	vertical rules.)
   \ncols=#1%  Get the number of columns desired.
   \begingroup%	 Start local parameter definitions.
   \let\ARGS=0%	 This is the key to the	whole thing; it	prevents \ARGS
%		 from being expanded in	the following \edef's.
   \edef\xtp{\widevline\ARGS\tabskip\tabskipglue%
   &\tstrut\ctr{\ARGS}}%  A 1-column preamble.
   \advance\ncols by -1%  One column has been generated; decrement the
%			  counter.
   \loop%  Append as many further columns as needed to the preamble.
      \ifnum\ncols>0 %
      \advance\ncols by	-1%
      \edef\xtp{\xtp&\vrule width\thinsize\ARGS&\ctr{\ARGS}}%
   \repeat
   \xdef\preamble{\xtp&\widevline\ARGS\tabskip0pt%
   \crnorm}%  Adds the last \vrule.
   \endgroup%  End of local parameters.
}%  End	of macro \makePREAMBLE.
\def\countROWS#1\into#2{%  This	counts the number of rows in #1	by
%			   looking for control sequences that end a row,
%			   e.g., \cr, \crthick,	etc., and puts the result
%			   into	count register #2.
   \let\countREGISTER=#2%
   \countREGISTER=0%
%  \out{In countROWS:  tokens are [\the#1]}%
   \expandafter\ROWcount\the#1\endcount%
}%
\def\ROWcount{%
   \afterassignment\subROWcount\let\next= %
}%
\def\subROWcount{%
%  \out{In subROWcount:	 next is [\meaning\next]}%  Debugging aid.
   \ifx\next\endcount %
      \let\next=\relax%
   \else%
      \ncase=0%
      \ifx\next\cr %
	 \global\advance\countREGISTER by 1%
	 \ncase=0%
      \fi%
      \ifx\next\endrow %
	 \global\advance\countREGISTER by 1%
	 \ncase=0%
      \fi%
      \ifx\next\crthick	%
	 \global\advance\countREGISTER by 1%
	 \ncase=0%
      \fi%
      \ifx\next\crnorule %
	 \global\advance\countREGISTER by 1%
	 \ncase=0%
      \fi%
      \ifx\next\header %
%     \out{In subROWcount:  next=header, ncase set=1}%
	 \ncase=1%
      \fi%
%     \out{In subROWcount:  ncase is [\the\ncase]}%
      \relax%
      \ifcase\ncase %
	 \let\next\ROWcount%
%	 \out{subROWcount---> ncase=\the\ncase}%
      \or %
	 \let\next\argROWskip%
%	 \out{subROWcount---> ncase=\the\ncase}%
      \else %
      \fi%
   \fi%
%  \out{subROWcount--->	NEXT=\meaning\next}%
   \next%
}%  End	of macro \subROWcount.
\def\counthdROWS#1\into#2{%
\dvr{10}%
   \let\countREGISTER=#2%
   \countREGISTER=0%
\dvr{11}%
%  \out{In counthdROWS:	 tokens	are [\the#1]}%
\dvr{13}%
   \expandafter\hdROWcount\the#1\endcount%
\dvr{12}%
}%
\def\hdROWcount{%
   \afterassignment\subhdROWcount\let\next= %
}%
\def\subhdROWcount{%
%\out{In subhdROWcount:	 next is [\meaning\next]}%
   \ifx\next\endcount %
      \let\next=\relax%
   \else%
      \ncase=0%
      \ifx\next\cr %
	 \global\advance\countREGISTER by 1%
	 \ncase=0%
      \fi%
      \ifx\next\endrow %
	 \global\advance\countREGISTER by 1%
	 \ncase=0%
      \fi%
      \ifx\next\crthick	%
	 \global\advance\countREGISTER by 1%
	 \ncase=0%
      \fi%
      \ifx\next\crnorule %
	 \global\advance\countREGISTER by 1%
	 \ncase=0%
      \fi%
      \ifx\next\header %
%\out{In subhdROWcount:	 next=header, ncase set=1}%
	 \ncase=1%
      \fi%
%\out{In subhdROWcount:	 ncase is [\the\ncase]}%
\relax%
      \ifcase\ncase %
	 \let\next\hdROWcount%
%\out{subhdROWcount--->	ncase=\the\ncase}%
      \or%
	 \let\next\arghdROWskip%
%\out{subhdROWcount--->	ncase=\the\ncase}%
      \else %
      \fi%
   \fi%
%\out{subhdROWcount--->	NEXT=\meaning\next}%
   \next%
}%
{\catcode`\|=13\letbartab
\gdef\countCOLS#1\into#2{%
%  \out{In countCOLS:  tokens are [\the#1]}
   \let\countREGISTER=#2%
   \global\countREGISTER=0%
   \global\multispancount=0%
   \global\firstrowtrue
   \expandafter\COLcount\the#1\endcount%
   \global\advance\countREGISTER by 3%
   \global\advance\countREGISTER by -\multispancount
%  \out{countCOLS-->[\the\countREGISTER]}
}%
\gdef\COLcount{%
   \afterassignment\subCOLcount\let\next= %
}%
{\catcode`\&=13%
\gdef\subCOLcount{%
%\out{In subCOLcount: next is [\meaning\next]}
   \ifx\next\endcount %
      \let\next=\relax%
   \else%
      \ncase=0%
      \iffirstrow
	 \ifx\next& %
	    \global\advance\countREGISTER by 2%
	    \ncase=0%
	 \fi%
	 \ifx\next\span	%
	    \global\advance\countREGISTER by 1%
	    \ncase=0%
	 \fi%
	 \ifx\next| %
	    \global\advance\countREGISTER by 2%
	    \ncase=0%
	 \fi
	 \ifx\next\|
	    \global\advance\countREGISTER by 2%
	    \ncase=0%
	 \fi
	 \ifx\next\multispan
	    \ncase=1%
	    \global\advance\multispancount by 1%
	 \fi
	 \ifx\next\header
	    \ncase=2%
	 \fi
	 \ifx\next\cr	    \global\firstrowfalse \fi
	 \ifx\next\endrow   \global\firstrowfalse \fi
	 \ifx\next\crthick  \global\firstrowfalse \fi
	 \ifx\next\crnorule \global\firstrowfalse \fi
      \fi%  End	of \iffirstrow.
\relax%\out{subCOL-->  ncase=[\the\ncase]}
% \out{subCOL-->  next=\meaning\next}
      \ifcase\ncase %
	 \let\next\COLcount%
      \or %
	 \let\next\spancount%
      \or %
	 \let\next\argCOLskip%
      \else %
      \fi %
   \fi%
%  \out{subCOL-->  countREGISTER=[\the\countREGISTER]}
   \next%
}%
\gdef\argROWskip#1{%
%  Deletes the next balanced, undelimited argument from	a
%		  token	list.
% \out{---> Entering argROWskip	<---}
% \out{In argROWskip:  deleted arg is [#1]}%
   \let\next\ROWcount \next%
}%  End	of macro \argskip.
\gdef\arghdROWskip#1{%
%  Deletes the next balanced, undelimited argument from	a
%		  token	list.
% \out{---> Entering arghdROWskip <---}
% \out{In arghdROWskip:	 deleted arg is	[#1]}%
   \let\next\ROWcount \next%
}%  End	of macro \arghdROWskip.
\gdef\argCOLskip#1{%
%  Deletes the next balanced, undelimited argument from	a
%		  token	list.
% \out{---> Entering argCOLskip	<---}
% \out{In argCOLskip:  deleted arg is [#1]}%
   \let\next\COLcount \next%
}%  End	of macro \argskip.
}%  End	of active &'s.
}%  End	of active |'s.
\def\spancount#1{%\out{spancount--->\meaning#1}
   \nspan=#1\multiply\nspan by 2\advance\nspan by -1%
   \global\advance \countREGISTER by \nspan
%  \out{number spancount--->\the\nspan;	\the\countREGISTER}
   \let\next\COLcount \next}%
\def\dvr#1{\relax}%
% \omit\hfil%
% \parindent=0pt\hsize=1.1in\valign{%
% \vfil#\vfil&\vfil#\vfil\cr\hfil\hbox{\ Added to\ }\hfil&%
% \hfil\hbox{\ empty events\ }\hfil\cr}\hfil%
\def\header#1{%
\dvr{1}{\let\cr=\@mpersand%
\hdtks={#1}%
%\out{In header:  hdtks=[\the\hdtks]}%
\counthdROWS\hdtks\into\hdrows%
\advance\hdrows	by 1%
\ifnum\hdrows=0	\hdrows=1 \fi%
%\out{In header:  Nhdrows=[\the\hdrows]}%
\dvr{5}\makehdPREAMBLE{\the\hdrows}%
%\out{In header:  headerpreamble=[\headerpreamble]}%
\dvr{6}\getHDdimen{#1}%
%\out{In header:  hdsize=[\the\hdsize]}%
%\striplastCR{#1}%
{\parindent=0pt\hsize=\hdsize{\let\ifmath0%
\xdef\next{\valign{\headerpreamble #1\crnorm}}}\dvr{7}\next\dvr{8}%
}%
}\dvr{2}}%  End	of macro \header.
\def\makehdPREAMBLE#1{%This macro generates the	necessary preamble for a
\dvr{3}%
%		       ruled table with	\ncols primary columns.
%		       (Primary	columns	means the number of columns NOT
%			counting those used for	vertical rules.
\hdrows=#1%  Get the number of columns desired.
{%  Start local	parameter definitions.
\let\headerARGS=0%
%  This	is the key to the whole	thing; it prevents \ARGS
\let\cr=\crnorm%
%		 from being expanded in	the following \edef's.
\edef\xtp{\vfil\hfil\hbox{\headerARGS}\hfil\vfil}%
\advance\hdrows	by -1%	One row	has been generated; decrement the
%			  counter.
\loop%	Append as many further rows as needed to the preamble.
\ifnum\hdrows>0%
\advance\hdrows	by -1%
\edef\xtp{\xtp&\vfil\hfil\hbox{\headerARGS}\hfil\vfil}%
\repeat%
\xdef\headerpreamble{\xtp\crcr}%
}%  End	of local parameters.
\dvr{4}}%  End of \makehdPREAMBLE.
\def\getHDdimen#1{%
%\out{In getHDdimen:  Arg 1=[#1]}%
\hdsize=0pt%
\getsize#1\cr\end\cr%
}%  End	of macro getHDdimen.
\def\getsize#1\cr{%
%\out{In getsize:  Arg 1=[#1]}%
%  Here	we have	to check arg#1 and see if the first token in #1	is an
%    \end; if so, we stop, else	we check the width of arg#1.
%  We recall that each arg#1 will be terminated	with a \cr token.
\endsizefalse\savetks={#1}%
%\out{In getsize:  the savetks = [\the\savetks]}%
\expandafter\lookend\the\savetks\cr%
%\out{In getsize:  ifendsize = [\meaning\ifendsize]}%
\relax \ifendsize \let\next\relax \else%
\setbox\hdbox=\hbox{#1}\newhdsize=1.0\wd\hdbox%
\ifdim\newhdsize>\hdsize \hdsize=\newhdsize \fi%
%\out{In getsize:  hdsize=[\the\hdsize]}%
%\out{In getsize:  newhdsize=[\the\newhdsize]}%
\let\next\getsize \fi%
\next%
}%
\def\lookend{\afterassignment\sublookend\let\looknext= }%
\def\sublookend{\relax%
%\out{In sublookend:  looknext = [\looknext]}%
\ifx\looknext\cr %
%\out{In sublooknext:  looknext=cr}%
\let\looknext\relax \else %
%\out{In sublooknext:  looknext/=cr}%
   \relax
   \ifx\looknext\end \global\endsizetrue \fi%
   \let\looknext=\lookend%
    \fi	\looknext%
}%
%
%  Allow the user to make his own names	for crthick, etc.
%
\def\tablelet#1{%
   \tableLETtokens=\expandafter{\the\tableLETtokens #1}%
}%
\catcode`\@=12%	 Change	@'s back to their normal category code.
%

%%%%%%%%%%%%%%%% MACRO ENDS %%%%%%%%%%%%%%%%%%%%%%

\centerline {\ }
\vskip 1.0in
\centerline {\bf {PATTERNS IN NON-LINEAR GRAVITATIONAL CLUSTERING:}}
\smallskip
\centerline {\bf {A NUMERICAL INVESTIGATION}}
\vskip 1.0in
\centerline {T.Padmanabhan \footnote*{Permanent address: IUCAA, Post Bag. 4, Ganeshkhind, Pune 411 007, India.  email: paddy@iucaa.ernet.in}, Renyue Cen, Jeremiah P. Ostriker, and F J Summers}
\centerline {\it Princeton University Observatory}
\centerline {\it Princeton, NJ 08544}
\centerline {email: cen@astro.princeton.edu}
\vskip 1.0in
\centerline{In Press of {\it The Astrophysical Journal}}
\vfill\eject

\centerline{\bf ABSTRACT}
\noindent
The nonlinear clustering of dark matter particles in an expanding universe
is usually studied by N-body simulations. One can gain some insight into
this complex problem if simple relations between physical
quantities in the linear and nonlinear regimes can be extracted
from the results of N-body simulations.
 Hamilton et al. (1991) and Nityananda and Padmanabhan (1994) 
 have  
made an attempt in this direction by relating  the mean relative
pair velocities to the mean correlation function in a useful manner. We investigate this relation and other closely related issues 
in detail for the case of six different power
spectra: power laws with spectral indexes $n=-2,-1$, cold dark matter (CDM),
and hot dark matter models with density parameter $\Omega=1$; CDM including a
cosmological constant ($\Lambda$) with $\Omega_{CDM}=0.4$,
$\Omega_{\Lambda}=0.6$; and $n=-1$ model
with $\Omega=0.1$. We find that: (i) Power law spectra lead to self-similar
evolution in an $\Omega=1$ universe. (ii) Stable clustering does not hold
in an $\Omega=1$ universe to the extent our simulations can ascertain. (iii)
Stable clustering is a better approximation in the case of  $\Omega<1$
universe in which structure formation freezes out at some low redshift. (iv) The
relation between dimensionless pair velocity and
the mean correlation function, $\bar\xi$, is only approximately independent
of the shape of the power spectrum. At the nonlinear end, the asymptotic
value of the dimensionless pair velocity decreases with increasing small scale power, because the stable clustering
assumption is not universally true.  (v) The relation
between the evolved $\bar\xi$ and the linear regime $\bar\xi$ is also not universal
but shows a weak spectrum dependence.  We present simple theoretical arguments for these conclusions. 
\vskip 2.cm
\noindent 
Cosmology: large-scale structure of Universe 
-- cosmology: theory
-- galaxies: clustering
-- numerical methods
\vfill\eject

\beginsection\centerline{1. INTRODUCTION}

Large-scale structures (like galaxies, clusters etc.)
are believed to have evolved primarily through the action
of gravitational force out of small initial density 
perturbations.
When these density perturbations are small, it is possible to study
their evolution using linear theory. But once the 
density contrast becomes comparable to unity, linear
perturbation theory breaks down and one must invoke
either analytic approximations or direct N-body simulations to study the growth of perturbations. 

Several people have performed  extensive numerical simulations, with varying
dynamic range and resolution, to study this problem.
While these simulations are of  value in making concrete predictions 
for specific models, they do not seem to have provided deep  insight into physics of non-linear gravitational dynamics. 
To obtain such an insight into  this complex problem, 
it is important to ask whether simple patterns in 
gravitational clustering can be extracted from 
the plethora of
simulations. If such patterns are found, they will
work as pointers for possible analytic approximations
and could even eliminate the need for extensive simulations in some specific cases. We shall study, in this paper, the possible existence of such simple patterns in the nonlinear gravitational clustering.

It would be  naive to hope for a simple picture 
to emerge in the full theory of structure formation, involving both gravitational
and hydrodynamic processes. 
Radiative coupling of baryons and 
photons is a multiscale process with  several 
dimensionless ratios, and it is unlikely that such
processes can be described by simple universal rules. 
On scales less than  a few Mpc such physical effects become
increasingly important.
The situation is quite different as regards 
the dark matter in the universe, often assumed to be composed of 
collisionless,
weakly interacting, massive particles. By  limiting our consideration to the  gravitational
interaction alone, one may hope to uncover some basic relationships.

Section 2 discusses some possible scaling relations and motivates the numerical simulations we have done. Section 3 describes the simulations and last section
presents the results and discussions.

\beginsection\centerline{2. PATTERNS IN NONLINEAR CLUSTERING}

For the case of interest,
we can describe the dark matter as a system of 
particles interacting via Newtonian gravity,
at scales which are small compared to the Hubble radius. 
The motion of particles in this case can be described by the equations
$$\ddot \x_i + {2 \dot a \over a } \dot \x_i = - {1 \over a^2} \nabla \phi; \quad \nabla^2 \phi = 4 \pi G \rho_b a^2 \delta, \eqn\qfrwpt $$
where $\x_i\lb t \rb$ 
is the comoving position of the i-th particle, $a\lb t \rb$ is the expansion factor, $\rho_b \lb t \rb$ is the background density in a matter dominated universe and $\phi$ is the gravitational potential due to the perturbed density 
$\delta \rho \equiv \rho - \rho_b \equiv \rho_b \delta $. All spatial derivatives are with respect to the comoving coordinates. In the fluid limit, 
the same system is described by the equations
$$\dot \delta + \nabla \cdot \left[ \lb 1 + \delta \rb \u \right] = 0; \quad \dot \u + \lb \u \cdot \nabla \rb \u + {2 \dot a \over a } \u = - {1 \over a^2 } \nabla \phi, \eqn\qfrwfl$$
where $\u \equiv \lb d \x /dt \rb$. For a universe with density parameter $\Omega = 1 $, we can introduce the time coordinate 
$\tau \equiv \ln \lb t / T \rb$  where $T$ is an arbitrary positive constant and a rescaled potential 
$\psi \equiv \lb 4 / 9 \rb H_0^{-2} \lb a \phi \rb $ with $H_0$ the Hubble
constant. Then equation \qfrwpt \ reduces to
$$ {d^2 \x_i \over d \tau^2 } + {1 \over 3 } {d \x_i \over d \tau} = - \nabla \psi;\quad \nabla^2 \psi = \delta, \eqn\qrdpt $$ 
while \qfrwfl \ becomes
$${\partial \delta \over \partial \tau } + \nabla \cdot \left[ \lb 1 + \delta \rb \p \right] = 0; \quad {\partial \p \over \partial \tau } + \lb \p \cdot \nabla \rb \p + {1 \over 3 } \p = - \nabla \psi, \eqn\qq $$
where $\bf p$ is  the velocity field corresponding to $\tau$ coordinate.
All reference to the background spacetime has completely disappeared in these equations. Hence these equations have no intrinsic scale. If we now further assume that the 
initial power spectrum of perturbations is scale invariant 
with $P(k)=Ak^n$, then we expect the dynamical
evolution of such a system to possess some universal characteristics.

To avoid misunderstanding, we stress several points. First, any particular realization of an initial power spectrum will have scales appearing in it randomly. When averaged over an ensemble of realizations with
the same $P \lb k \rb$, these scales will disappear.
Second, the use of a grid to generate initial conditions will introduce a
scale of order the mean interparticle distance. Such scales are erased by evolution,
except in the most underdense void regions.  Third, softening in the
gravitational force introduces a minimum resolution scale which cannot be
avoided. We shall, however, concentrate on scales which are larger than the softening scale to obtain reliable conclusions.
Fourth, requirements of convergence, $\lb n > -3 \rb$, and the generation of a
$k^4$ tail due to discreteness effects restrict the range of $n$ to $-3 < n < 4 $. If equations  \qfrwfl\ are studied perturbatively, spurious divergences due to long wavelength modes can arise for $-3 < n < -1 $. The long wavelength contribution to bulk velocity, for example, has a divergent contribution for $n = -2 $ when handled perturbatively. The nonperturbative argument given in the paragraph above clearly shows that these divergences are spurious. Thus endpoint
effects should not be important for $-3<n<4$ and thus no physical scales enter the problem for $\Omega=1$.

One immediate consequence of the above analysis is that  a simple scaling 
relation {\it must} exist in all simulations involving
dark matter with power law spectrum, in an
$\Omega$ = 1 universe. Since there is no a priori
scale in the problem, 
the natural length at any time is 
$x_{{\rm NL}}(a)$, 
the comoving scale which is going nonlinear at the epoch $a$ . 
Thus all dimensionless 
physical parameters must be expressible as a universal
function of $(x/x_{{\rm NL}})$.
For example, we expect the 
correlation function $\xi (x,a)$ to have the form 
$\xi (x,a) = f_n [x/x_{{\rm NL}} (a)]$. This implies that 
the evolution is self-similar, and $\xi$  is completely 
specified by a single function $f_n$ which depends on the index $n$.

Further progress can be achieved if we assume that clustering 
at the extreme nonlinear end is ``stable'' (Peebles 1980).
In such a case, $\xi (a,x)$ must have the form
$\xi (a,x) = a^3 F(ax)$ in the extreme nonlinear limit, because the number density  of
objects in collapsed regions is assumed constant, while the average
density of the universe decreases as $a^{-3}$.
This assumption is equivalent to the postulate that highly  
overdense, virialized objects maintain their identities, i.e.,
merging and fragmentation either do not occur or
occur at strictly balancing rates.
The assumption of self-similarity requires that
the correlation function at the 
linear end evolves as $\xi_L \propto a^2 x^{-(n+3)}$. 
Combining this result with the stable
clustering postulate, it is easy to show that
$$\xi (a,x) \propto a^3 (ax)^{-\gamma} \propto a^{6/(n+5)} x^{-3(n+3)/(n+5)}, \eqn\qfive$$
in the extreme nonlinear end.
In other words, the index of the nonlinear correlation
function $\gamma$ should be
related to the index of the linear power spectrum $n$ by
$$\gamma = {3(n+3) \over (n+5)}, \eqn\qsix $$
if the stable clustering ansatz were true.
Therefore, given the correlation function in the linear theory,
$\xi_L$, it is possible to determine the correlation 
function in the nonlinear theory up to a constant factor provided
we assume the validity of stable clustering. 
In section 4 we examine
the quantitative accuracy of the above two assumptions. 

Equation \qsix , 
however, is not very useful in practice.  
To begin with, it works only in the {\it extreme} 
nonlinear regime and one is often interested in the
intermediate scales. Second, it does not provide
any clue to the amplitude of the nonlinear correlation
function in terms of the linear one. Third, it is
based on the assumption that the power spectrum is
strictly scale free. 
And finally, it rests on the ad hoc assumption of stable clustering.

Attempts have been made to generalize this idea 
by using an effective index $n_{{\rm eff}}$ as a 
function of scale, defined as,
$$n_{{\rm eff}} \equiv {d\ln P \over d\ln k}.\eqn\qq$$
Nityananda \& Padmanabhan (1994, hereafter NP) found that a simple
implementation of scaling using $n_{eff}$ does not work satisfactorily in that a smooth transition
between linear and non-linear regimes is not obtained.
It is important,
therefore, to look for more sophisticated - yet manageable - rules which will provide information about the nonlinear theory in terms of the linear theory. 

One such attempt was made in NP along the following lines: They assumed that 
the mean relative pair velocity of particles $v_p (a,x)$ 
can be expressed in the form 
$$v_p (a,x) = - \dot a x h [ \bar \xi (a,x)], \eqn\qrntp$$
where $\bar\xi$ is the average correlation function within a sphere of radius $x$ and $h$ is a universal function of its argument. Given this ansatz, NP show that it is possible to relate the mean correlation function $\bar\xi$ to the mean linear correlation function $\bar \xi_L$ by the relation
$$\bar \xi_L (a,l) = \exp \lb {2\over 3} \int^{\bar\xi (a,x)} {d\nu \over h(\nu) (1+\nu)} \rb, \eqn\qbarel$$
with $l \equiv x [1+ \bar \xi (a,x)]^{1/3}$. 
Earlier work by Hamilton et al. (1991)
suggested that such a relationship seems to be borne out by the
numerical simulations. In other words, numerical 
simulations seem to suggest that the function $h(\bar \xi)$ 
is universal; it is independent of the epoch {\it and}
the power spectrum used in the simulations. This function seems to behave as follows: $h \lb \bar \xi \rb \approx \lb 2/3 \rb \bar \xi$ for $\bar \xi \ll 1 $, reaches a maximum of about 2 at $\bar \xi \approx 20 $ and decreases to about unity at large $\bar \xi$. For a wide class of power law spectra and spectra similar to that of
cold dark matter (CDM), the relationship in \qfive\ can also be expressed as 

$$\bar \xi (a, x )=\cases{\bar \xi_L (a,l) &(for\ $\bar \xi_L<1.2, \, \bar \xi<1.2$)\cr
0.7[\bar \xi_L  (a,l)]^3  &(for\ $1.2<\bar \xi_L<6.5, \, 1.2<\bar \xi<195$)\cr
11.7[\bar \xi_L  (a,l)]^{3/2} &(for\ $6.5 <\bar\xi_L, \, 195<\bar \xi$)\cr}.
\eqn\qbagh$$
This fitting function is due to Bagla and Padmanabhan (1993) and
captures the essence of a more complicated 
fit given in Hamilton et al (1991).

Unlike the self-similarity argument mentioned earlier, this  relation cannot be independent of the power spectrum in the strict sense. To see this, consider a power spectrum  with a sharp maximum at $ x \approx L$ and very little small scale power. [This is similar to the HDM spectrum]. The first scales which go nonlinear will correspond to clusters with size $x \approx L$. The evolution of the first nonlinear structures in such a model could be well approximated by collapse of a spherical top hat (STH), since there is very little small scale power. For an STH model, one can easily show that $h$ is a monotonically increasing function of the density contrast $\delta \approx \bar \xi$ and $h \approx \bar \xi^{1/2}$ as $\bar \xi \rightarrow \infty$ (see
the discussion in section 4 below). This suggests that $h$ will increase with $\bar \xi$ for a good range of $\bar \xi$ in the HDM-like models until significant amount of small scale power develops. One can reach the same conclusion from \qbagh \ as well: If there is very little small scale power in linear theory, then $\bar \xi$ will have to rise far more steeply with $\bar \xi_L$ (in the nonlinear end) than suggested by \qbagh.

The study of such an extreme example suggests that there might exist a weak spectrum dependence in the $h \lb \bar \xi \rb$ relation even in more moderate cases. Broadly speaking, we expect $h$ to be lower at a given $\bar \xi$ (in the nonlinear regime with $\bar \xi > 10 $ or so ) as we add more small scale power.          If stable clustering is invoked, then it is possible to motivate this conclusion along the following lines: 

Consider a spherical region of initial radius $r_i$ and overdensity $\nu \sigma$ where $\sigma = \sigma_0 r_i^{-(n + 3)/2}$ is the variance of the gaussian density fluid. In a spherical model, this region will expand to a maximum radius of about $( r_i / \nu \sigma )$ and virialise to a final radius $r \equiv \lambda (r_i / \nu \sigma ) = ( \lambda / \nu \sigma_0 ) r_i ^{n + 5 \over 2} $ where $\lambda \approx 0.5$.
We shall assume that the coorelation function at the nonlinear end 
$\xi_{NL} (r)$
is contributed by such virialised objects and can be computed as

$$ 1 + \xi (r) \approx \xi (r) = \langle \lb {r_i \over r } \rb^3 \rangle. \eqn\qrirav $$
This assumption is equivalent to the stable clustering hypothesis. From $r = ( \lambda / \nu \sigma_0) r_i^{(n + 5)/2}$,
it follows that

$$\xi (r) = \lb {\sigma_0 \over \lambda} \rb ^{6/(n + 5 )} r ^{-{3 (n + 3 ) \over (n + 5 ) }} \left \langle \nu ^{6/(n + 5 )} \right \rangle. \eqn\qsinl $$
Assuming that $\nu $ is a gaussian variable, we get

$$\langle \nu^{6/(n + 5 )} \rangle = {1 \over \sqrt{2 \pi} } 2 ^{{ 1 -n \over 2n + 10 }} \Gamma \lb {n + 11 \over 2n + 10 } \rb. \eqn\qnuav $$
Equations \ \qsinl \ and \ \qnuav \ show that $\xi (r) \propto \xi^{3/2}_L [l]$ with $l \approx r \xi^{1/3}$; however, the proportionality constant has a weak n-dependence.

The averaging in \qrirav\ can be made more
sophisticated by using a weightage proportional 
to $r^m_i$. In that case, we still obtain the same $r$ dependence but the gamma
function becomes

$$c = \sqrt{\pi}{\Gamma \lb {n + 11 + 2m \over 2n + 10} \rb \over \Gamma \lb {n+5+2m \over 2n+10 }\rb}.\eqn\qq $$
This reduces to the above result when $m = 0;$ recently, Mo et al. (1995) have suggested a model based on $m = 3 $, which -- of course -- leads to similar conclusions.

While the above arguments suggest that $h[\bar\xi]$ will have a weak spectrum dependence, they depend crucially on the assumption of stable clustering. The $\xi \lb r \rb$ can be modelled by virialised clusters only if stable clustering is valid. This statement is independent of the actual averaging process used in \ \qsinl\ . Because of this reason, arguments like the above
ones -- and the $n$ dependence obtained from them -- are unreliable. {\it If the system does not obey stable clustering criterion, $h [\xi]$ may still show an n-dependence but it cannot be obtained from the above argument.} On the other hand, a
theoretical explanation for the approximate universality (and the deviations from it) will go a long way in providing an analytic description of the nonlinear evolution of gravitating particles. It is, therefore, important to test this ansatz as precisely as possible in as diverse cases as possible. In particular,
one must ask: (i) Do the relations $h = h \lb \bar \xi \rb $ and equations \ \qbagh \ have a weak spectrum dependence for pure power law spectra in $\Omega = 1 $ universe? (ii) How are the relations modified if  the spectrum is not a pure
power law? and (iii) What happens when the  background universe is not described by a matter dominated,
$\Omega =1$ model?
We present the results of such
a numerical investigation in this paper.

\bigskip 
\centerline{\bf 3. NUMERICAL MODELING}
\smallskip
%\beginsection\centerline{3. NUMERICAL MODELING}

Two kinds of N-body simulations are used
to simulate the evolution of the particles.
The first one uses
a standard Particles-Mesh code (PM, \cf\ Hockney \& Eastwood 1981;
Efstathiou \etal\ 1985)
with the staggered mesh scheme (see, e.g., Park 1990; Cen 1992).
We use $240^3=10^{7.1}$ particles on a $720^3$ mesh with
a simulation box size of $128h^{-1}$Mpc giving a spatial
resolution of $0.18h^{-1}$Mpc and
mass resolution of $4.2\times 10^{10}\Omega h^{-1}M_\odot$.
Gravitational force is calculated using an FFT technique coupled
with the cloud-in-cell scheme to calculate the density
as well as to assign the gravitational force to the positions
of particles.

The second kind of simulations uses the P3MG3A code (Brieu,
Summers, \& Ostriker 1995). This code is based on the Particle-Particle-Particle-Mesh 
(P$^3$M) scheme (Efstathiou and Eastwood 1981; Efstathiou \etal 1985)
utilizing a special gravity processor called
GRAPE 3A (Okumura \etal\ 1993) to compute the PP part.
Each simulation uses $128^3=10^{6.3}$ particles on a $256^3$ mesh with
a simulation box size of $128h^{-1}$Mpc.
The Plummer softening length is set to be $1/20$ of the mesh cell
size, giving a nominal spatial resolution of $0.025h^{-1}$Mpc and
a mass resolution of $2.8\times 10^{11}\Omega h^{-1}M_\odot$.

Six different models are considered and are listed in Table 1.  In the
table, $L_{box}$ refers to the one dimensional size of the computational
cube, $l_{res}$ lists the nominal resolution scale, and $\sigma_8$ gives
the linear value of the rms fluctuations in the mass on a sphere 
of radius 8~$h^{-1}$ Mpc at redshift zero.
We adopt the transfer functions of
Bardeen \etal (1986) 
for the standard CDM and HDM models.
The power spectrum of the 
CDM+$\Lambda$ model is from Cen, Gnedin, \& Ostriker (1993).
The initial density fields for each simulation
are generated assuming Gaussian fluctuations and
the initial velocity field is  obtained using the 
Zel'dovich approximation.

\beginsection\centerline{4. RESULTS AND DISCUSSION}

The above simulations were utilized to both examine the assumptions and explore
the consequences of the arguments in \S2. The important points to address are
the self-similarity of scale free simulations, the validity of stable
clustering, the universal nature of the relation between the dimensionless
pair velocity and the mean correlation function, and the dependencies on
spectral index. We discuss each of these features below:

\beginsection\centerline{\it 4.1 Self-similar evolution}

We begin by verifying the self-similar nature of the evolution for power law spectra by testing whether $\xi \lb a, z \rb $ can be expressed as $\xi \left[ x /x_{nl} \lb a \rb \right]$. This question was investigated earlier by 
Efstathiou et al. (1988)
who obtained results which were somewhat different   from ours.
We have now repeated this analysis with better resolution and range for
$n=-2,-1$ and we find that {\it self-similarity is borne out}
in our simulations as expected. 
For the case $n=-1$ the fit is excellent.
For $n=-2$ it is less good but, we believe, adequate. 
Figures (1a,b) show
the two-point correlation function as function
of $x/x_{nl}$ (where $x_{nl}$ is the scale corresponding to
$\xi=1$ at three different redshifts $z=0.0,0.5,1.0$
for $n=-2,-1$ models. Note that the curves fall on top of each other.
Also shown (as two dashed lines in each figure)
are asymptotic slopes at the extreme nonlinear and linear ends.
This result is expected 
based on the theoretical arguments given in section 2,  
but serves as a good check on the 
accuracy of simulations. In addition, it clarifies certain
doubts raised in Efstathiou et al. (1988) regarding self-similarity 
for the case $n =-2$.
On all the figures we only plot the points which correspond
to scales (separations) larger than 2.5 cells in the PM simulations
and twice the softening lengths in the P$^3$M simulations,
which represent the true resolutions of two types of simulations.

It may be noted that in the case of $n = -1 $, the linear slope extends to the 
quasi-linear regime as well, up to about $\bar \xi \approx 10$. This fact
can be understood from equation \qbagh. In the quasi-linear regime, this equation predicts the evolution
$$ \bar \xi\propto a^{{6\over (n+4)}}x^{-{3(n+3)\over(n+4)}}, \eqn \qq $$
so that $\bar \xi \propto a^2 x^{-2}$ for $n = -1 $ in the quasi-linear regime as well. In other words, the linear and quasi-linear evolution has the same $ a $ and $x$ dependence for $n = -1 $. (See Bagla and Padmanabhan, 1995 for a more detailed discussion of this point.)

For comparison, 
We also show in Figures (1c,d,e,f) the corresponding 
results for the other four models (listed in Table 1):
SCDM (1c), HDM (1d), LCDM (1e), and open $P_k=k^{-1}$ (1f) models,
respectively. 
%item 10
All these models have a characteristic scale and
the low-$\Omega$ models have that feature even at the highest
redshift probed. 
%item 10
This should lead to deviations from self-similarity
in figures 1 c - e. Such a deviation is apparent 
in the case of HDM (which has the strongest scale dependence),
somewhat less apparent but noticeable in the case
of CDM + $\lambda $ and CDM (in which the scale 
dependence is weaker) and ambiguous in the case of open
$n = -1 $. This is essentially due to the fact 
mentioned above: $n = -1 $ is special in the sense 
that linear evolution holds even for the quasi-linear regime.
Hence the deviations are somewhat offset in this case.

In a truly self-similar evolution, we also expect 
velocities of particles to be a function of 
$\left[ x/x_{nl} \lb a \rb \right]$.
To test this we studied
the behavior of the ratio $v_{3d}(a,x)/[\dot a x_{nonlinear}(a)]$ 
of the 3-dimensional velocity dispersion to
our unit of velocity.
Figures (2a,b,c,d,e,f)
show the results for the different cases.
It is seen that self-similarity (and the deviations from it) 
has the same pattern for velocity dispersion as for correlation functions:
Figures (2a,b) show time invariant behavior but (2c,d,e,f) do not. Note that
Figure 2f shows the non-self-similar behavior of the open $n=-1$ model that is
hidden in Figure 1f.
Note also the somewhat surprising result
that in all cases presented,
the velocity difference between 
two particles is a rather weak function of their separation
within $10^{\pm 1.5}$ of the nonlinear scale, especially for
the self-similar cases.

\medskip
\centerline{\it 4.2 Relative pair velocity, `universality' and stable clustering}
\smallskip

\noindent  According to the arguments presented in section 2, we expect curve $h(\bar\xi)$ to vary slightly but systematically as more small scale power is added. 
We indeed find  such a behavior. 

Figures (3a,b) show the behavior of $h$ for $n=-2$ and
$n=-1$ at three different redshifts $z=1,0.5,0$.
In the nonlinear end, the curves
for $ n=-1$ is lower than those for $ n=-2. $
That is, as we add more small scale power, 
the relative pair velocity decreases at nonlinear scales. 
This behavior
is further borne out in the case of CDM (see Figure 3c) and HDM (see Figure 3d).
In the redshift range considered, CDM model has an effective index which
is between n=-1 and n=-2 and HDM has less small scale
power than $ n=-2.$
The curve for CDM  falls in between those for  $ n=-1 $ and $ n=-2 $ 
and that of HDM is above the one for $n=-2$.

The behavior of $h  (\bar \xi) $ for the HDM model allows 
a simple test of the reasoning presented in section 2. 
To do this properly let us consider the
evolution of a spherical top hat model 
in a background universe with some $\Omega$ along the lines of 
Gunn and Gott (1972).
When $\Omega \not = 1 $, we have $a \lb t \rb \propto \lb 1 - \Omega^{-1} \rb $  where $\Omega $ now refers to the instantaneous value at time $t$. 
The quantity $h $ can be estimated in the STH model by the 
relation $\lb 1 + h \rb = ( \dot R / R )( a / \dot a )$,
where $R \lb t \rb $ is the radius of the spherical region;
similarly we may approximate $\bar \xi $ by the 
density contrast $\delta $. Both these approximations 
are valid for $\bar \xi \ga 1 $ and should be 
interpreted as some average value. From the STH model 
it is easy to show that (with $\bar \xi \equiv \delta $),
$$1 + \bar \xi = { \lb \mu - \sin \mu \rb^2 \over \lb 1 - \cos \mu \rb^3 } \quad {8 \lb \Omega -1 \rb^3 \over \Omega^3 \left[ \cos^{-1} \lb {2 \over \Omega }-1 \rb - {2 \over \Omega } \sqrt { \Omega -1 } \right]^2 }, \eqn\qpom $$
and
$$ 1 + h = {\lb \mu - \sin \mu \rb\sin \mu   \over \lb 1 - \cos \mu \rb^2 }\quad  {2 \lb \Omega - 1 \rb^{3/2} \over \Omega \left[ \cos^{-1} \lb {2 \over \Omega } -1 \rb - {2 \over \Omega } \sqrt {\Omega -1 } \right] }, \eqn\qhom $$
where $\mu$ is the phase angle describing the evolution of the STH. It follows that
$$ {\lb 1 + h \rb^2 \over 1 + \bar \xi }  = {\Omega \over 2 } \quad
{\sin^2 \mu \over \lb 1 - \cos \mu \rb}. \eqn\qnice $$
This is valid for all values of $\Omega$; equations \ \qpom \ and \qhom \ can be analytically continued for $\Omega \leq 1 $. 
A  nonlinear structure  which is undergoing free-fall collapse, we will have $\mu \approx 2 \pi - \epsilon $ with small $\epsilon $ and $\bar \xi \gg 1, h^2 \gg 1. $ In this limit equation \ \qnice becomes
$$h \cong \Omega^{1/2}_{eff} \bar\xi^{1/2}. \eqn\qq $$
The subscript ``eff'' is a reminder that we should use an effective $\Omega $ which exists around most collapsing structures at the epoch of interest. 
[We are assuming that the shperical top hat region can be modelled as being located inside a larger region, which - when smoothed over the relevant scale -
behaves like a universe with an effective $\Omega_{eff}$.]
Though it is difficult to model this quantity, two features are apparent: (1) At a given epoch, $h \propto \bar\xi^{1/2} $ in the nonlinear phase. (2) As clustering produces underdense regions more and more collapsing structures will find themselves in $\Omega_{eff} < 1 $ regions. As time goes on $\Omega_{eff}$ should decrease, and $h$ should decrease at a given $\bar\xi$.

Figure 3d  shows that both these expectations are borne out. The dashed lines at the nonlinear end has the slope of 0.5 which fits quite well to the simulation results. At a given $\bar \xi$, the $h$ decreases as time goes on; at $z \approx 0 $, we find $\Omega^{1/2}_{eff} \approx 0.3 $. 

In the case of Figures (3a,b) the key point to 
note is the lack of stable clustering.  In both the cases,
$h$ does {\it not } approach unity asymptotically. 
This fact is more easily discernible using $h$ than from
mean correlation function.
The value of $h$ at a given $\bar \xi $ decreases as we add more small scale power, as expected. Note that lack of stable clustering invalidates the argument based on \qsinl .
Also note that these results disagree with the results of simulations discussed in Mo et al. (1995). They claim $\xi (a, x ) \propto [ \xi_L (a, l ) ]^{3/2}$ asymptotically, which requires $h = 1 $ for large $\xi$. Our simulations show that $h \not= 1 $ asymptotically.

The models described above all have $\Omega=1$. 
To check the dependence on the 
background scale, we considered 
the behavior of $h[\xi]$ in the following cases:
(i) a flat CDM + $\Lambda$ model with $\Omega_{CDM} =0.4$ and $\Omega_\Lambda=0.6$
(model e) and (ii) an open model with $\Omega=0.1$  
and $P_k=Ak^{-1}$ (model f ).
These models have specific background scales and 
we find that it is reflected in $h[\xi]$ 
[see Figures (3e,f)]. 
The deviation in the linear end is
understandable since 
$$h \lb a, x \rb \equiv - {v_{\rm pair} \over \dot a x } = {2 \over 3 } \lb {a \dot b \over \dot a b } \rb \bar \xi \lb a, x \rb. \eqn\qq $$
where $b \lb t \rb$ is the growing solution to the linear
perturbation equation. For reasonable  $\Omega $, we can approximate $(a \dot b / \dot a b ) $ by $\left[ \Omega \lb z \rb \right]^{0.6}$. This formula is used usually at $z = 0$; it is a reasonable approximation at moderate $z$, if the instantaneous value of $\Omega$ is used. So, in the linear end $h \propto [\Omega \lb z \rb]^{0.6}$ at a given $\bar \xi$ and this is indeed seen in figure 3 (e, f). 
At the nonlinear end, $h$ seems  to approach unity within the limits of our 
resolution. This suggests that 
models  with $\Omega < 1 $ obey the hypothesis of stable
clustering to a better  approximation than models with $\Omega = 1 $. This is understandable since structures freeze out in the $\Omega < 1 $ model around $z \approx \Omega^{-1}-2$; in contrast, significant merging takes place in $\Omega = 1 $ universe at all epochs. In the intermediate scales, no clear pattern emerges in these models.

\medskip
\centerline{\it 4.3 Relation between $\bar\xi$ and $\bar\xi_L$}
\smallskip

The original ansatz can also be presented as a relationship between
$\bar\xi(a,x)$ and $\bar\xi_L(a,l)$. In the case of strict universality, these two approaches are related by equation \qbarel. However, when deviations from universality are present, it is worthwhile to examine both independently.
Figures (4a,b,c,d,e,f) show the results for different cases. 
These figures relate the nonlinear
correlation function at a scale $x$ to linear correlation function at a scale
$l$ where $l=x(1+\bar\xi)^{1/3}$.
The relation between $l$ and $x$ is independent of the value of
$\Omega$  and can be derived
by the method of characteristics used by NP. 
The broken lines in these figures have 
slopes 1 and 1.5 as suggested by equation \qbagh. 
We find that the deviations from our ansatz, seen in $h$, 
give rise to corresponding deviations in these figures. 
A spectral dependence  is
evident in Figures 4a and 4b, as well as in the slight shift for the CDM case,
Figure 4c .This dependence seems to be in the same sense as reported by Mo et al, 1995.
The deviations from stable clustering follows the 
pattern discussed earlier, except for Figure 4f. When $\Omega \not = 1$, we have $\xi_L = b^2 (t)x^{-(n + 3 ) }$ and $\xi = a^3 F (ax) $; in that case, stable clustering assumption does not translate into a simple scaling of the form $\xi \propto \xi^{3/2}_L $.  By and large, these 
figures reinforce the conclusions arrived at earlier.

\medskip
\centerline{\it 4.4 Overall Conclusions}
\smallskip

1) As expected, (but contrary to some earlier work),
we find that if the input
$P_k$ has a power law dependence on $k$,
then the only scale that is relevant is the scale
at which mass fluctuations are nonlinear,
$x_{NL}$,
which induces a break in the input (linear) power spectrum.
Under these circumstances all quantities
that we have measured show a self-similar evolution, i.e.,
they are only  functions of scaled radius $x/x_{NL}$.

2) The ``stable clustering" assumption is not true in general,
presumably because, in the case of $\Omega=1$, there is steady
merging of structures so that isolated bound objects do not survive
intact, violating the key ansatz
behind the stable clustering hypothesis.
When $\Omega < 1$, since structure
freezes out at low redshift,
the stable clustering model is a better approximation than if
$\Omega=1$.

3) The relation between the dimensionless pair
velocity $h(a,x)=-v_{pair}/Hr$ and the
mean correlation function $\bar\xi(a,x)$ is not universal
because of the failure of the stable clustering assumption although
within the range of cosmologically plausible models
the relation $h(\bar\xi)$ shows sufficiently uniform
behavior to be a useful quantity.

This work was done while one of the authors (T.P) visited the Department of Astrophysical Sciences, Princeton during Sept-Nov, 94. T.P thanks 
Princeton for the warm hospitality. He also thanks the Smithsonian 
Institution which has supported his travel to USA under the Indo-US
Exchange Program.
The authors thank R. Nityananda for discussions. 
It is a pleasure to acknowledge the help of NCSA 
for allowing us to use their Convex-3880 supercomputer.
This research is supported in part by NASA grant
NAGW-2448, NSF grant AST91-08103 and NSF HPCC grant ASC-9318185.
RC and JPO would like to thank the hospitality of ITP during their stay
and the financial support from ITP
through the NSF grant PHY94-07194.
\vfill\eject

\centerline {\bf REFERENCES}
\smallskip
\refset
Bardeen, J.M., Bond, J.R., Kaiser, M., and Szalay, A.S. 1986, ApJ, 304, 15
\smallskip
\refset
Brieu, P.P., Summers, F.J., \& Ostriker, J.P. 1995, ApJ, submitted
\smallskip
\refset
Bagla, J.S. and Padmanabhan, T. 1993, IUCAA preprint 22/93; based on the lecture given at VI th IAU Asia-Pacific Regional Meeting on Astronomy, Aug,93; to appear in Jour. Astrophys. Astronomy.
\smallskip
\refset
Bagla, J.S. and Padmanabhan, T. 1995, Evolution of gravitational potential
in the quasilinear and nonlinear regimes, IUCAA preprint 8/95;astro-ph 9503077.
\smallskip
\refset
Cen, R. 1992, ApJS, 78, 341
\smallskip
\refset
Cen, R., Gnedin, N.Y., \& Ostriker, J.P. 1993, ApJ, 417, 387
\smallskip
\refset
Efstathiou, G., \& Eastwood, J.W. 1981, MNRAS, 194, 503
\smallskip
\refset
Efstathiou, G., Davis, M., Frenk, C.S., \& White, S.D.M. 1985, ApJS, 57, 241
\smallskip
\refset
Efstathiou, G., Frenk, C.S., White, S.D.M., \& Davis, M. 1988, MNRAS, 235, 715
\smallskip
\refset
Hamilton A.J.S., Kumar, P., Lu E., \& Mathews, A. 1991, ApJ, 374, L1
\smallskip
\refset
Hockney, R.W., \& Eastwood, J.W. 1981, ``Computer Simulations
Using Particles", McGraw-Hill, New York.
\smallskip
\refset
Gunn, J.E., \& Gott, J.R., III 1972, ApJ, 176, 1
\smallskip
\refset
Mo, H. J., Jain, B., \& White, S. D. M. 1995, MNRAS, submitted
\smallskip
\refset
Nityananda, R., \& Padmanabhan, T. 1994, MNRAS, 271, 976
\smallskip
\refset
Okumura, S. K., et al. 1993, PASJ, 45, 329
\smallskip
\refset
Park, C.B. 1990, MNRAS, 242, 59
\smallskip
\refset
Peebles, P.J.E. 1980, ``The Large-Scale Structure 
of the Universe" (Princeton: Princeton University Press)
\smallskip
\refset
\vfill\eject

\centerline{\bf FIGURE CAPTIONS}

\bigskip
\medskip
\hsize=5.25truein
\hoffset=2.45truecm
\item{Fig. 1--} 
Figures (1a,b,c,d,e,f) show
the two-point correlation function as function
of $r/r_{nl}$ (where $r_{nl}$ is the scale corresponding to
$\xi=1$) at three different redshifts 
[$z=0,0.5,1$ with (solid, dotted, dashed) curves, respectively;
this order will be maintained in subsequent figures] for
Models (a,b,c,d,e,f) listed in Table 1.
Also shown as two dashed lines in each figure
are asymptotic slopes at the extreme nonlinear and linear ends. Note that
figure 1 (a,b), which are for powerlaw spectra in $\Omega=1$ universe, show
the expected self similar behaviour while the other figures do not. 

\item{Fig. 2--} 
Figures (2a,b,c,d,e,f) show
the ratio $v_{3d}(a,x)/[\dot a x_{nonlinear}(a)]$ 
between
the 3-dimensional pairwise velocity dispersion to our velocity unit,
as a function of $r/r_{nl}$ 
at three different redshifts ($z=0,0.5,1$) for 
Models (a,b,c,d,e,f). Figure 2 (a,b), 
which are for powerlaw spectra in $\Omega=1$ universe, show
the expected self similar behaviour while the other figures do not. 

\item{Fig. 3--}
Figures (3a,b,c,d,e,f) shows $h$ as a function of $\bar\xi$
at three different redshifts 
[$z=0,0.5,1$ with (solid dots, open circles, open squares), respectively;
this order will be maintained in subsequent figures] 
for Models (a,b,c,d,e,f). 
The errorbars are shown for 1$\sigma$ statistical uncertainties.
In the nonlinear end the $h$ is lower for a given
$\bar\xi$ if there is more small scale power. The value of $h$ in the 
asymptotic limit depends on the spectrum and does not always go to unity;
this shows that stable clustering is not a valid assumption in general.

\item{Fig. 4--} 
Figures (4a,b,c,d,e,f) show $\bar\xi$ as a function of $\bar\xi_L$
at three different redshifts ($z=0,0.5,1$) for 
Models (a,b,c,d,e,f).

\vfill\eject

\centerline {{\bf Table 1.} Summary of the computed models}
\medskip
\begintable
\hfill  Run \hfill|
\hfill  Model \hfill|
\hfill  Code  \hfill|
\hfill  $\Omega$ \hfill|
\hfill  $\Lambda$ \hfill|
\hfill $L_{box}h ^{-1} Mpc$ \hfill|
\hfill $l_{res}h ^{-1} Mpc$ \hfill|
\hfill  $\sigma_8$  \hfill\cr
\hfill  a  \hfill|
\hfill  $P_k=Ak^{-2}$  \hfill|
\hfill  PM  \hfill|
\hfill  $1$  \hfill|
\hfill  $0$  \hfill|
\hfill  $128$  \hfill|
\hfill  $0.18$  \hfill|
\hfill  $1.05$  \hfill\cr
\hfill  b  \hfill|
\hfill  $P_k=Ak^{-1}$  \hfill|
\hfill  P$^3$M  \hfill|
\hfill  $1$  \hfill|
\hfill  $0$  \hfill|
\hfill  $128$  \hfill|
\hfill  $0.025$  \hfill|
\hfill  $1.05$  \hfill\cr
\hfill  c  \hfill|
\hfill  SCDM  \hfill|
\hfill  P$^3$M  \hfill|
\hfill  $1$  \hfill|
\hfill  $0$  \hfill|
\hfill  $128$  \hfill|
\hfill  $0.025$  \hfill|
\hfill  $1.05$  \hfill\cr
\hfill  d  \hfill|
\hfill  HDM  \hfill|
\hfill  PM  \hfill|
\hfill  $1$  \hfill|
\hfill  $0$  \hfill|
\hfill  $128$  \hfill|
\hfill  $0.18$  \hfill|
\hfill  $1.05$  \hfill\cr
\hfill  e  \hfill|
\hfill  LCDM  \hfill|
\hfill  P$^3$M  \hfill|
\hfill  $0.4$  \hfill|
\hfill  $0.6$  \hfill|
\hfill  $128$  \hfill|
\hfill  $0.025$  \hfill|
\hfill  $0.79$  \hfill\cr
\hfill  f  \hfill|
\hfill Open $P_k=Ak^{-1}$ \hfill|
\hfill  PM  \hfill|
\hfill  $0.1$  \hfill|
\hfill  $0$  \hfill|
\hfill  $128$  \hfill|
\hfill  $0.18$  \hfill|
\hfill  $1.05$  \hfill
\endtable
\end

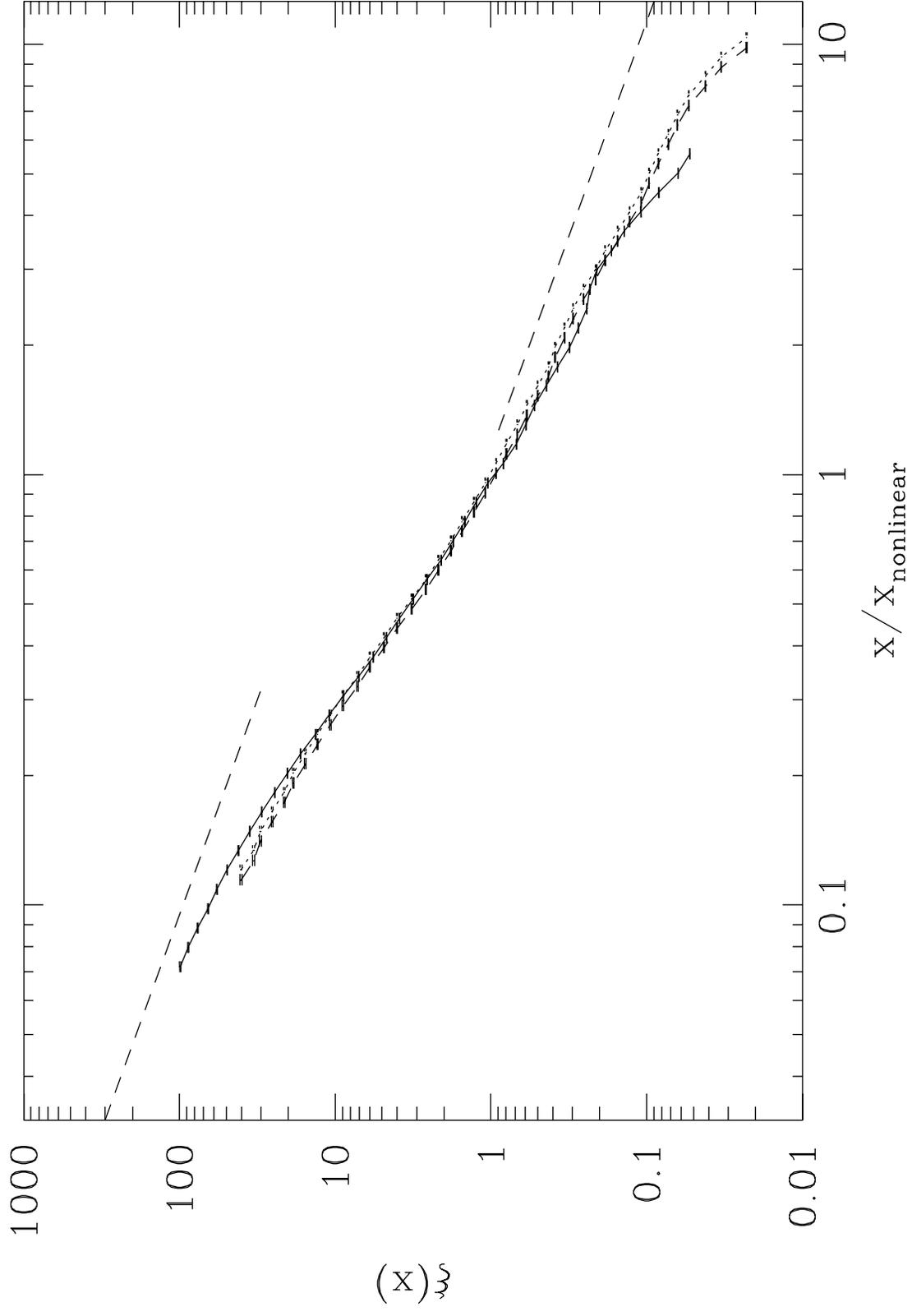
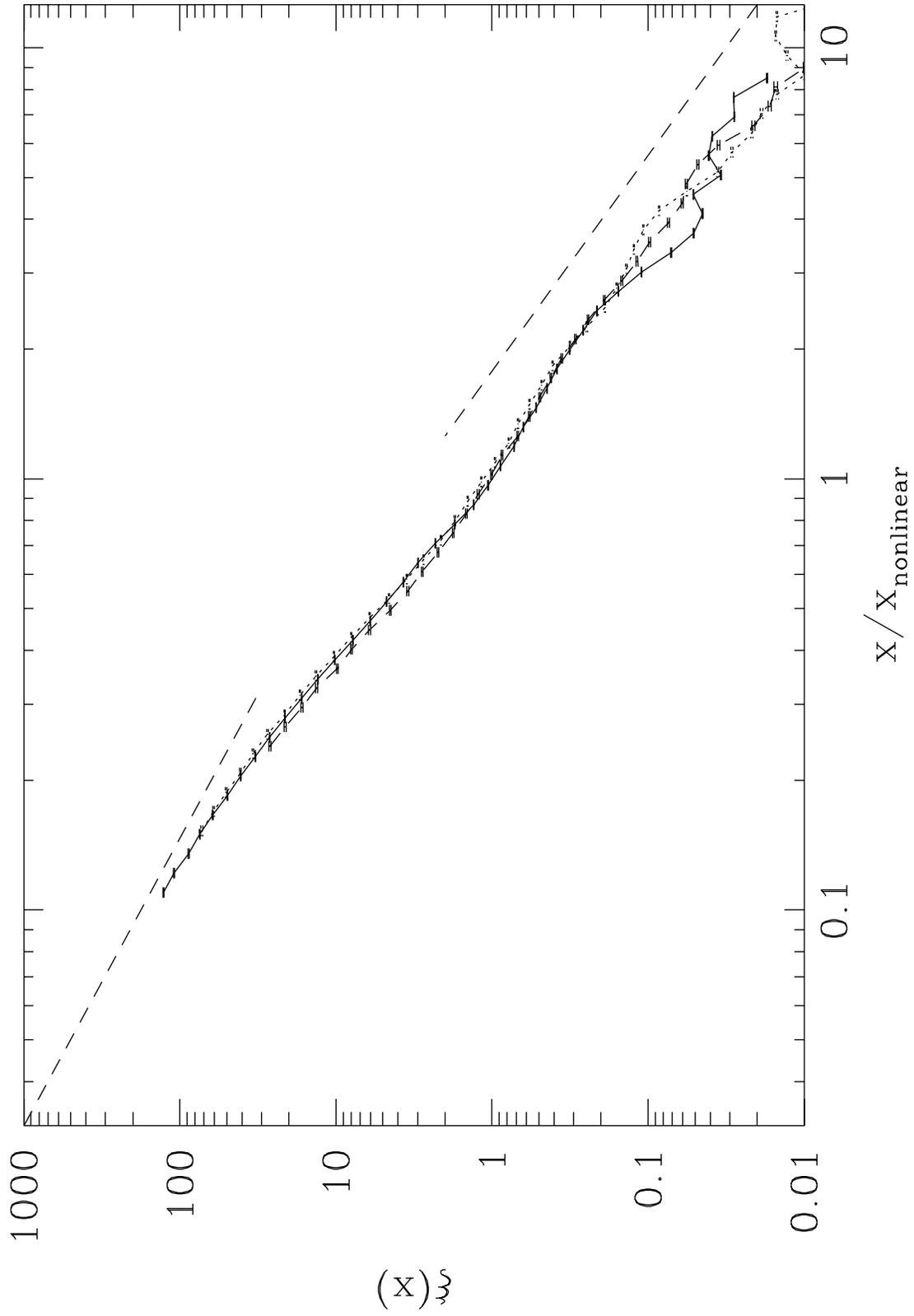
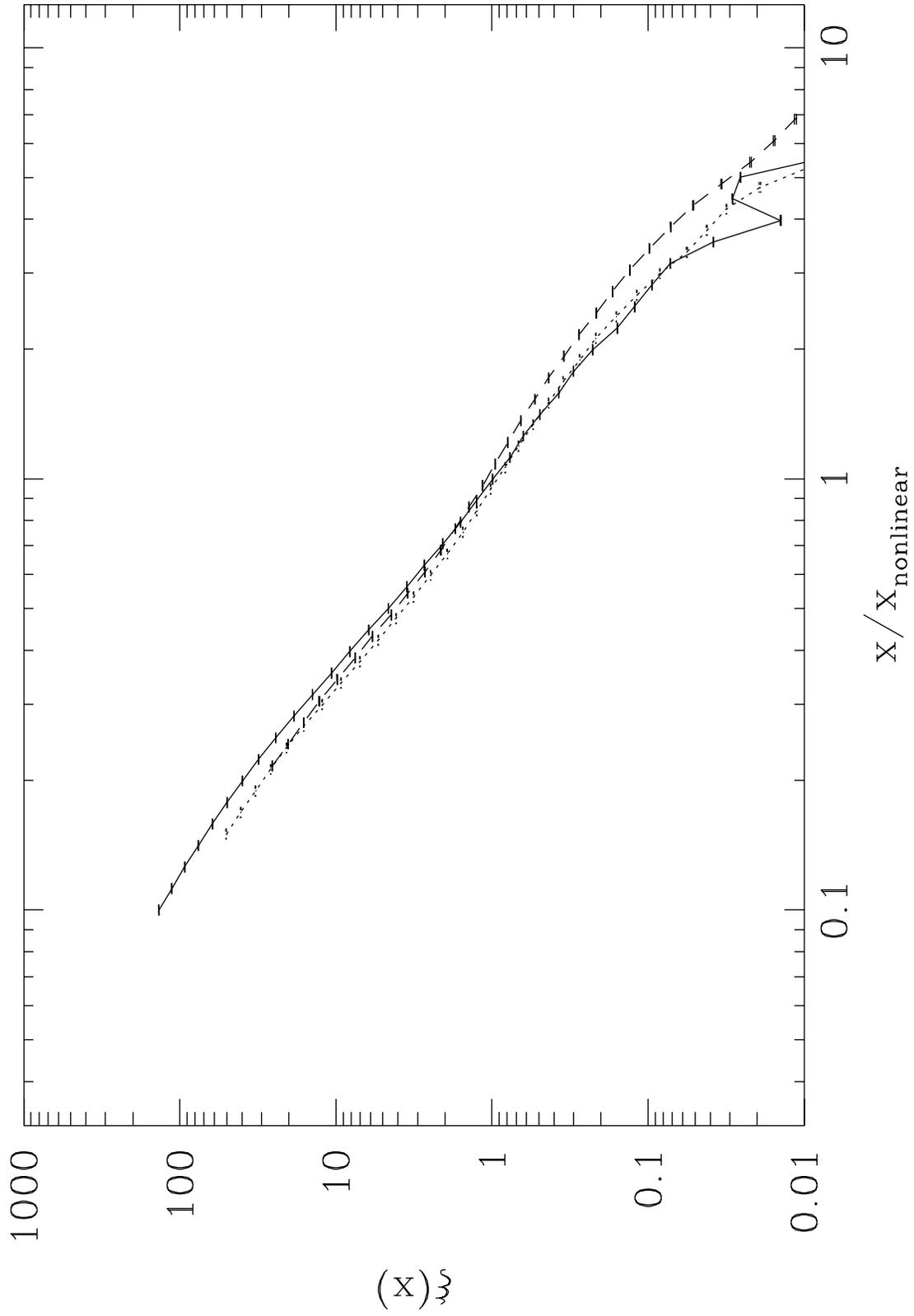
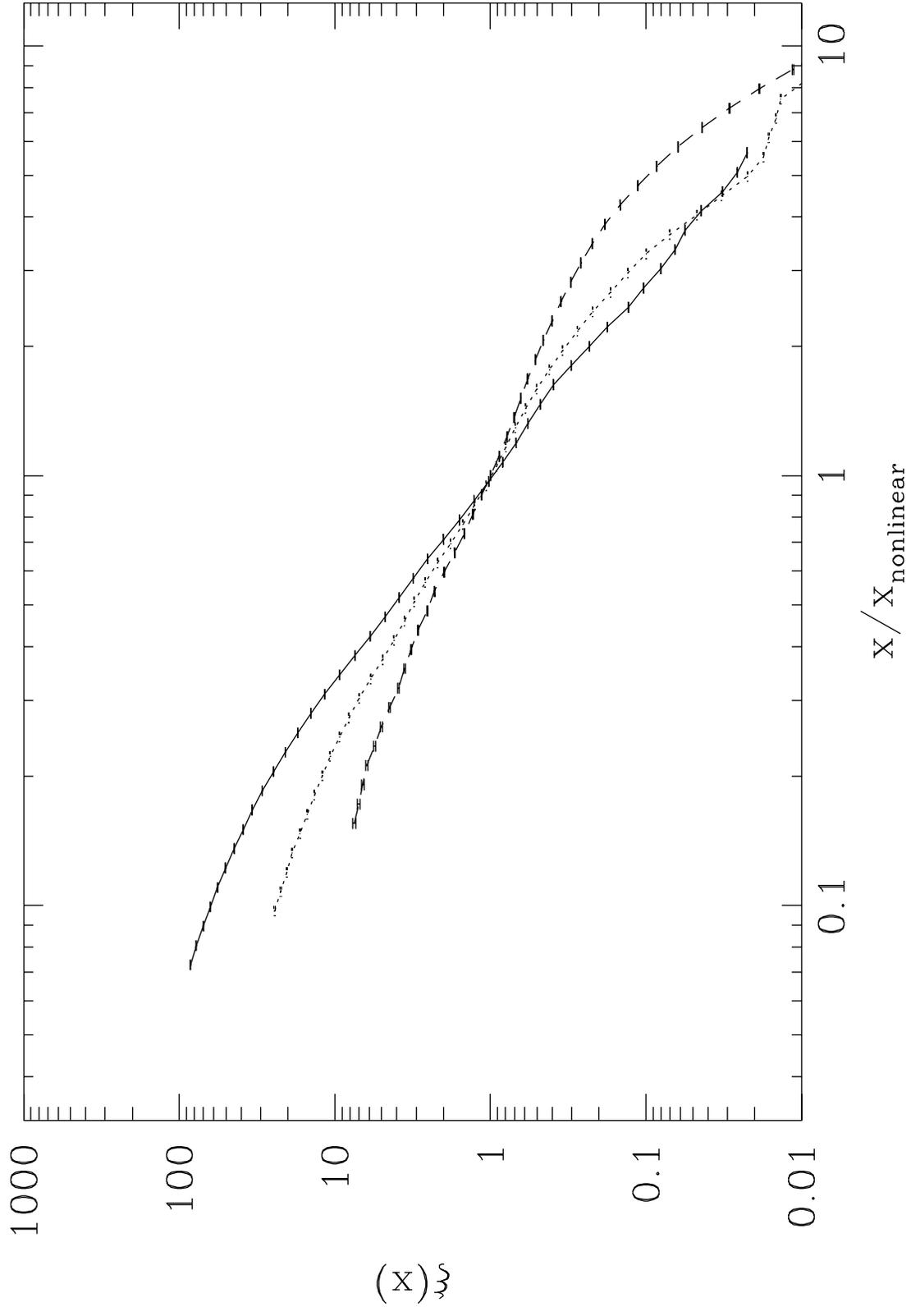
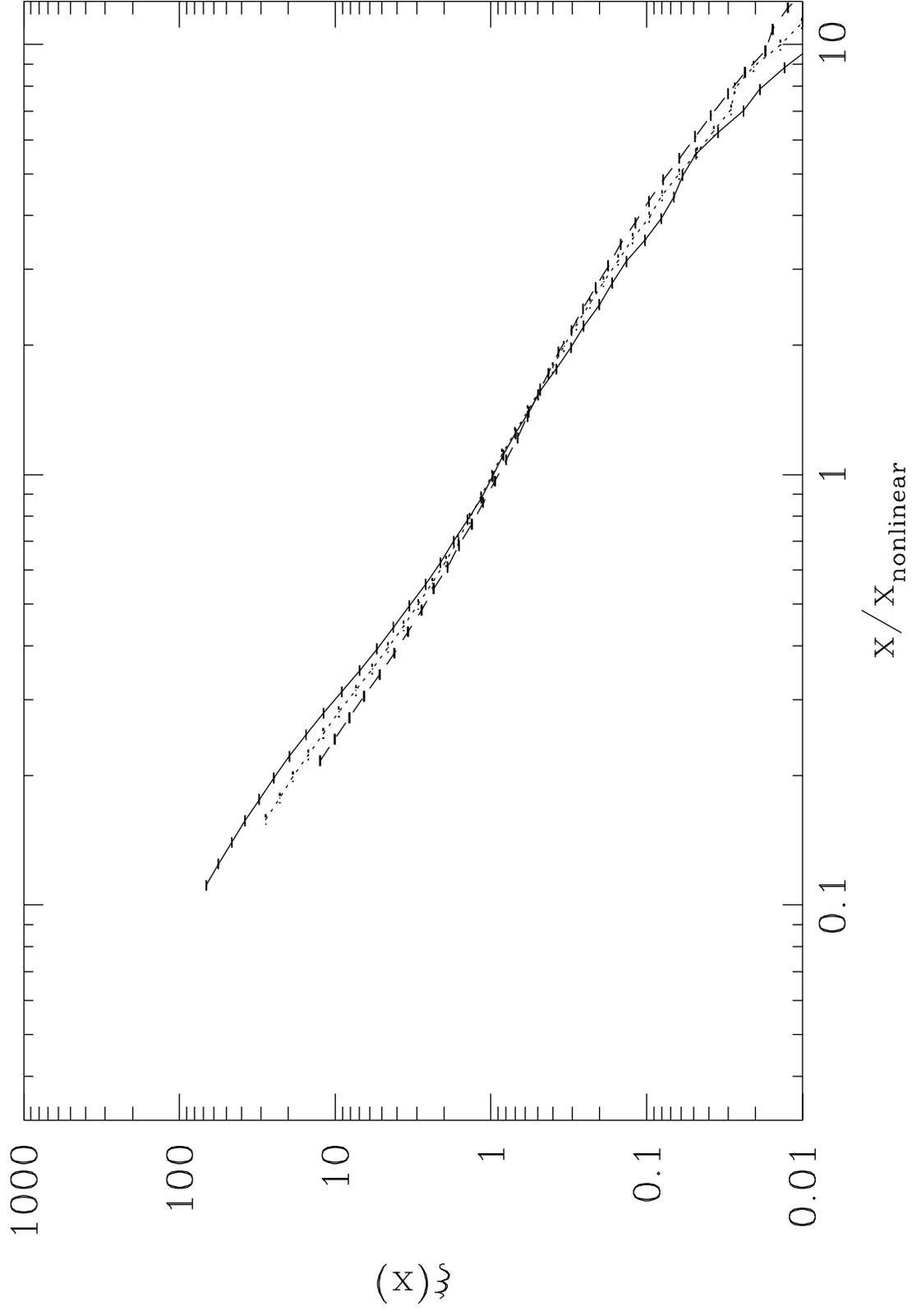
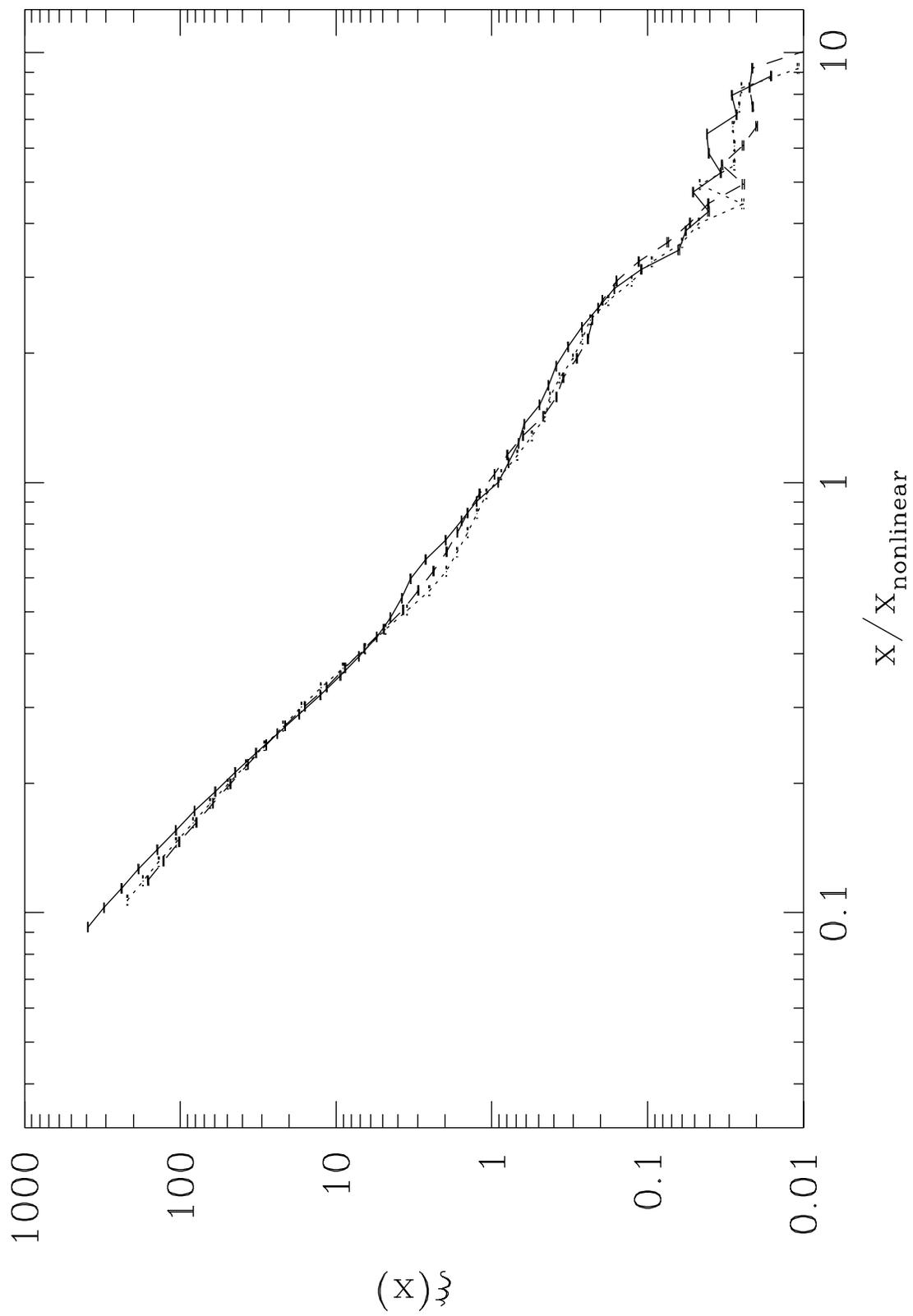
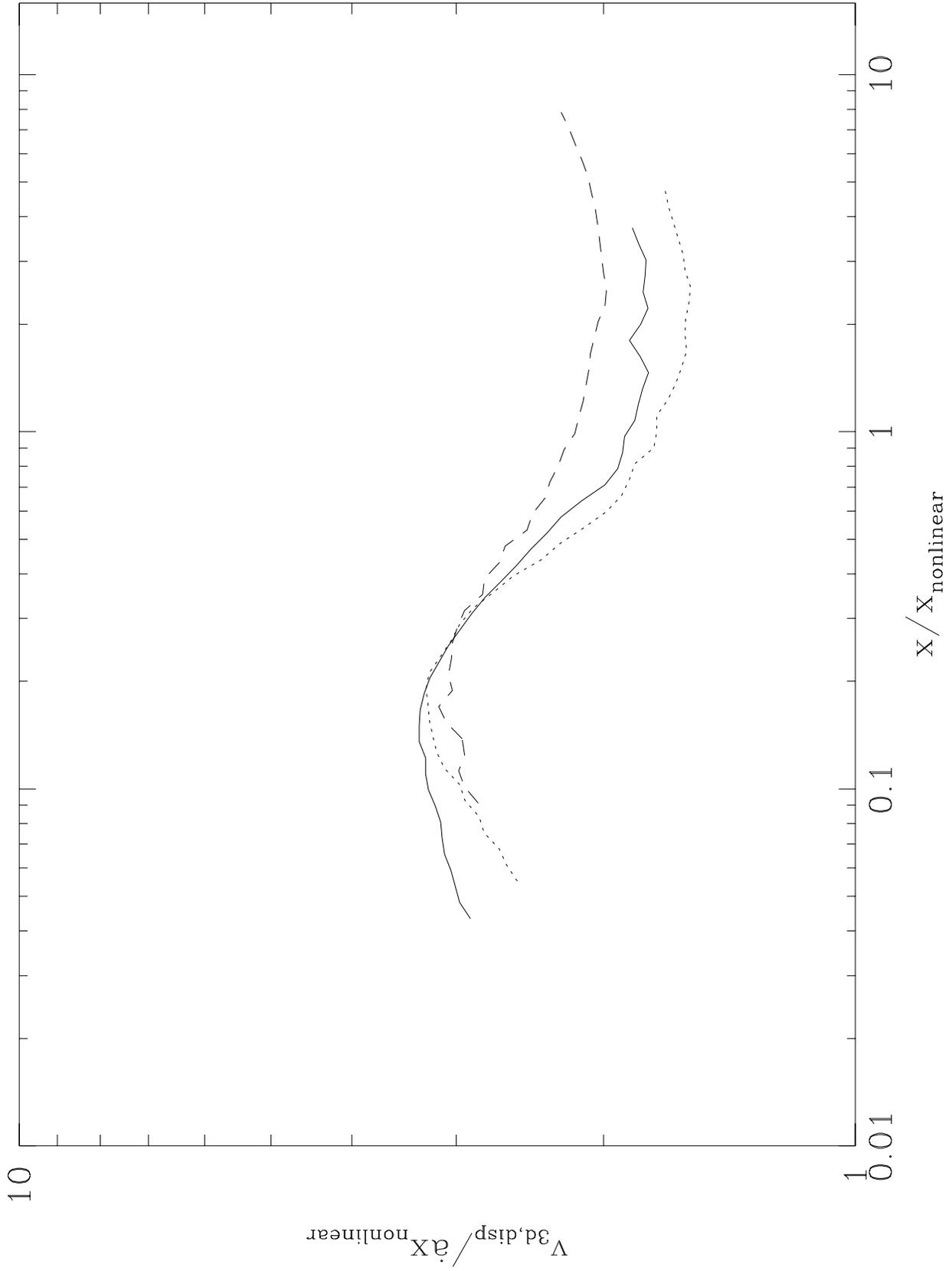
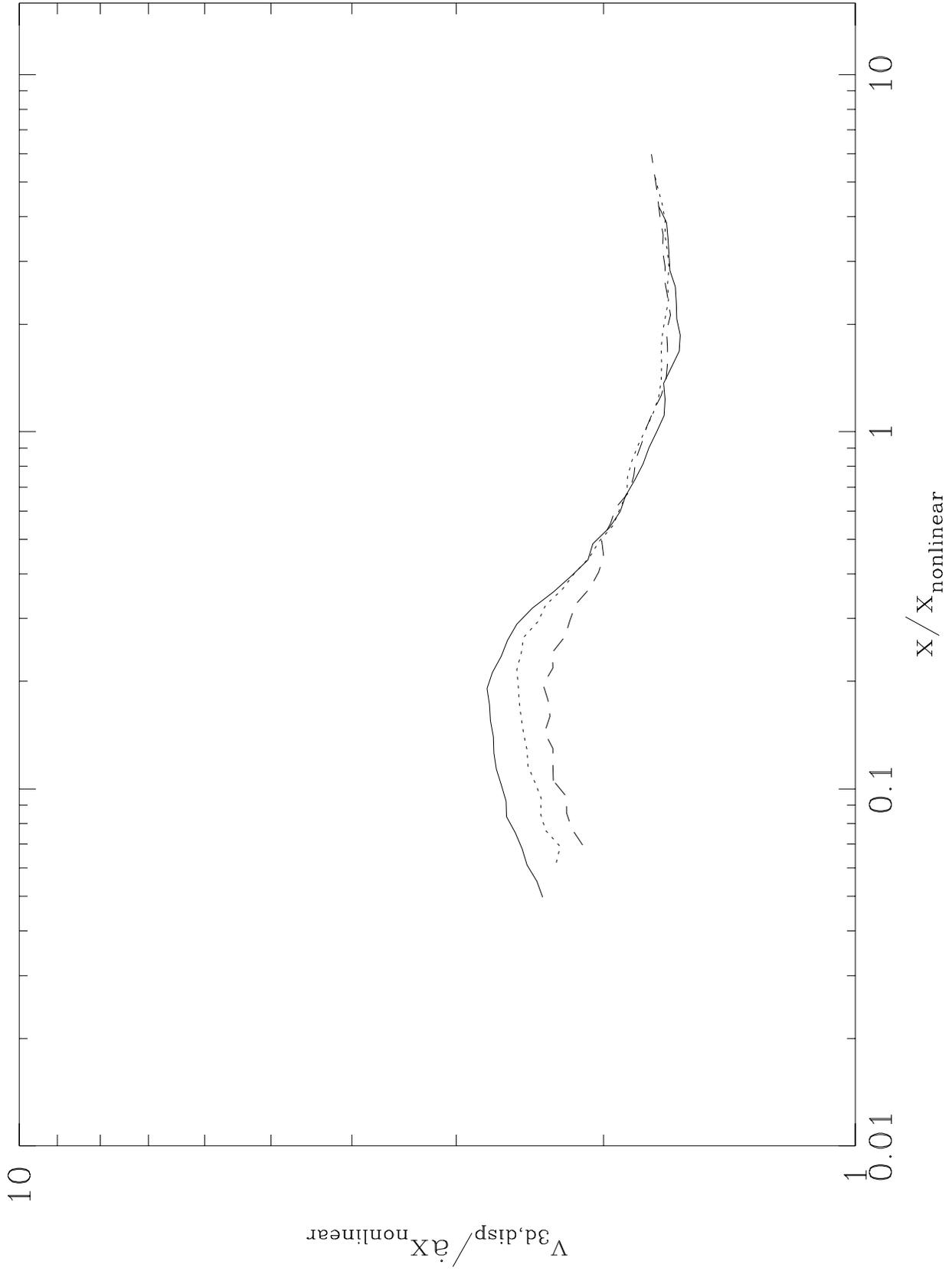
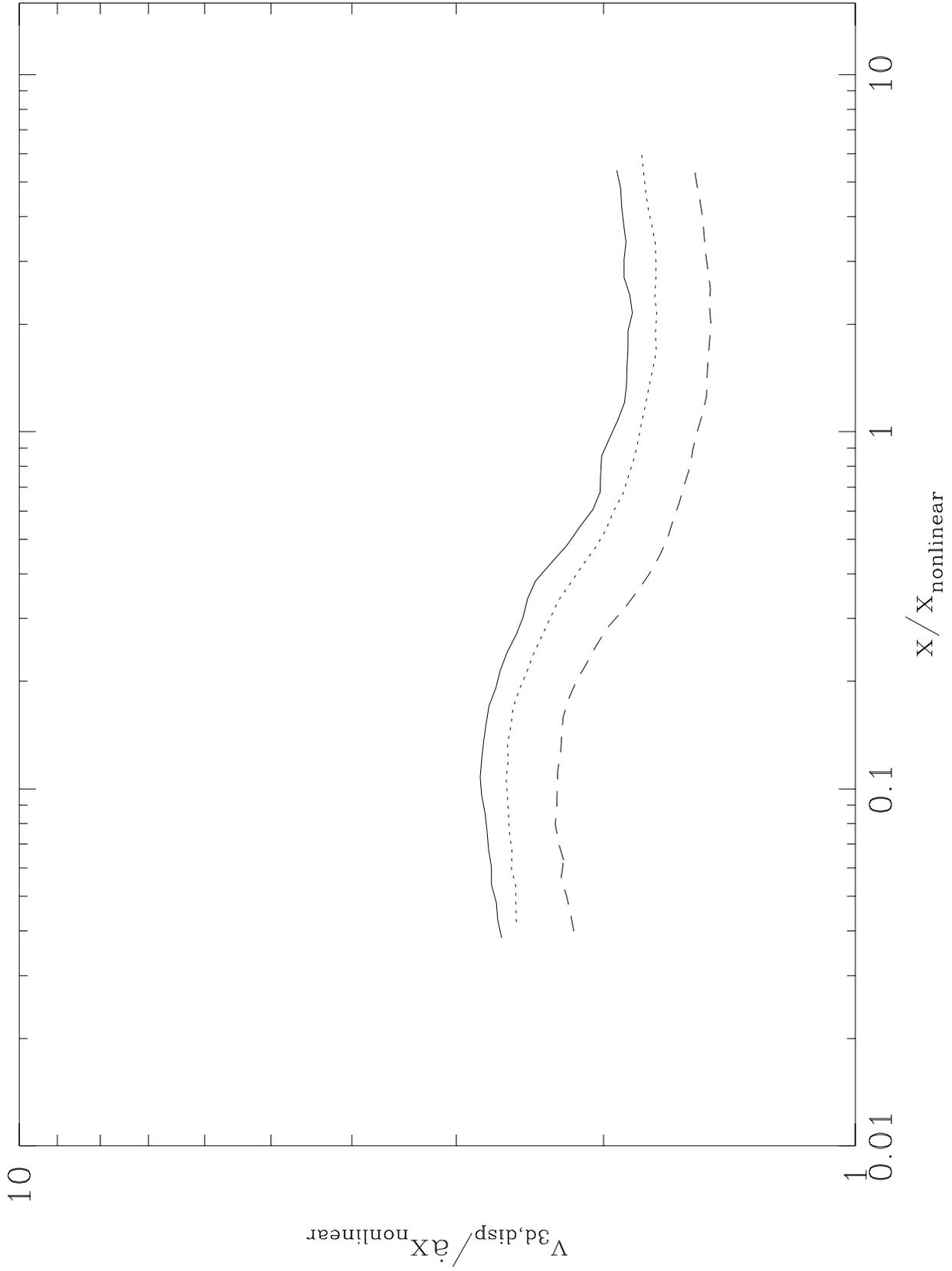
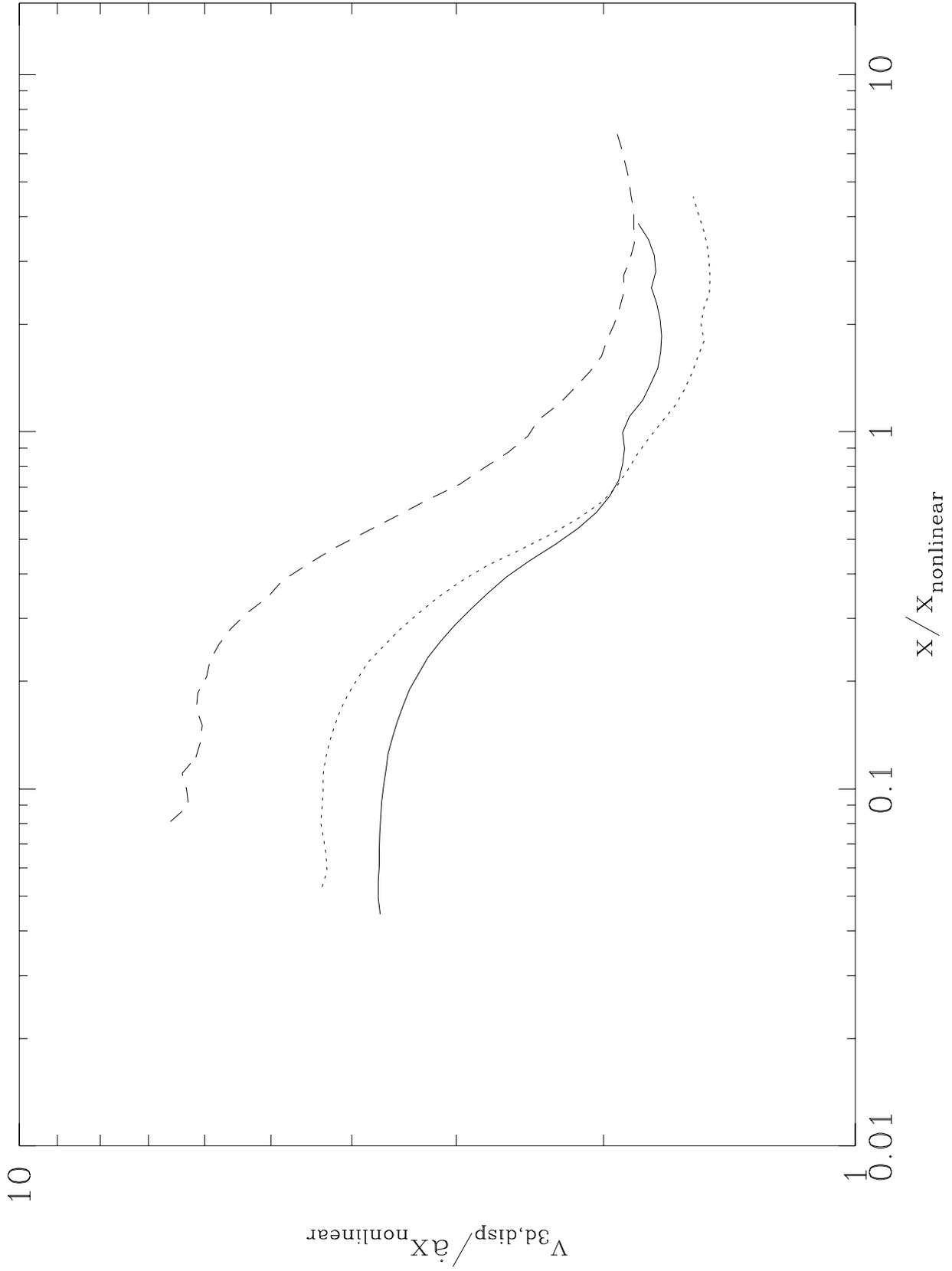
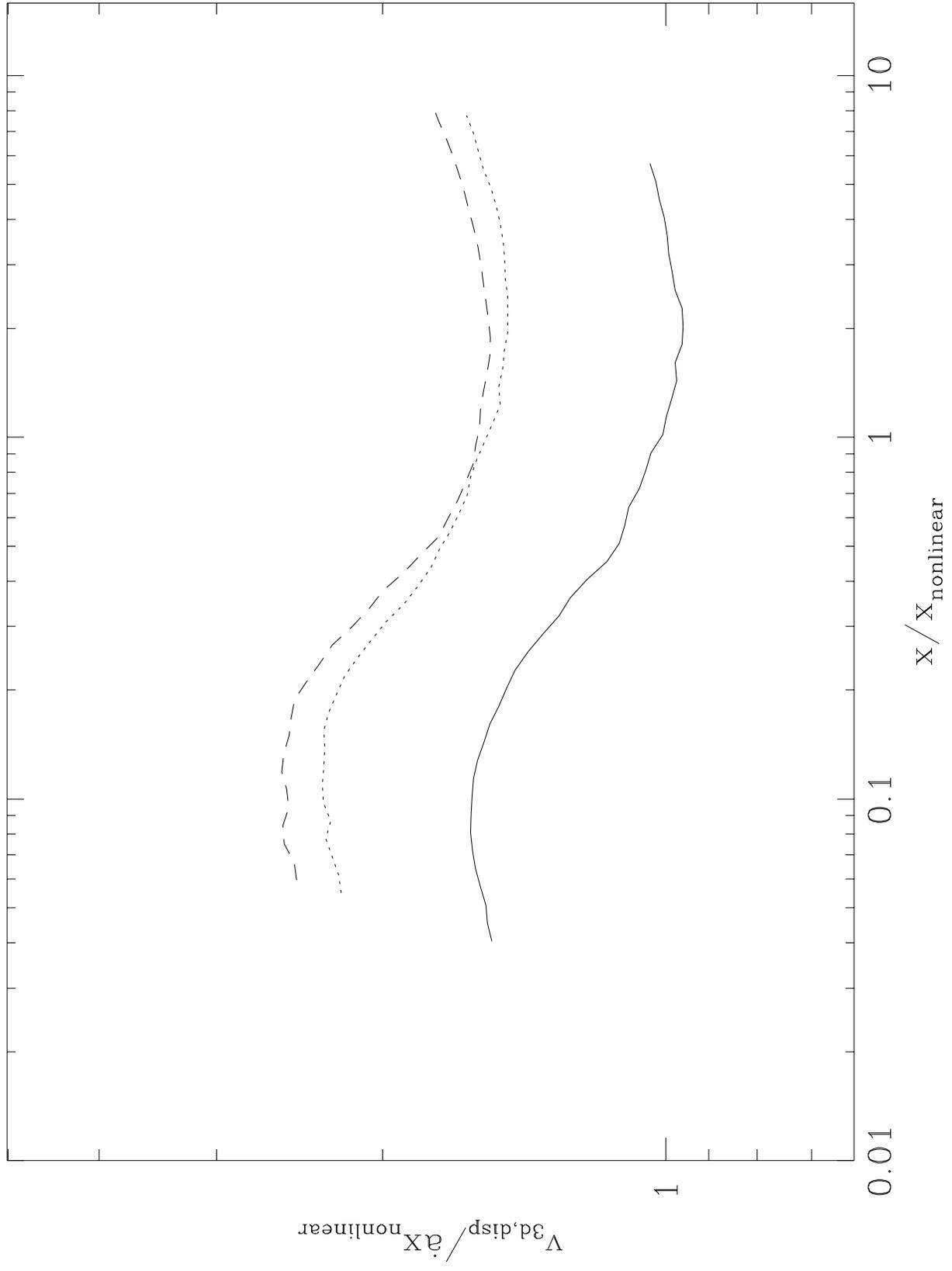
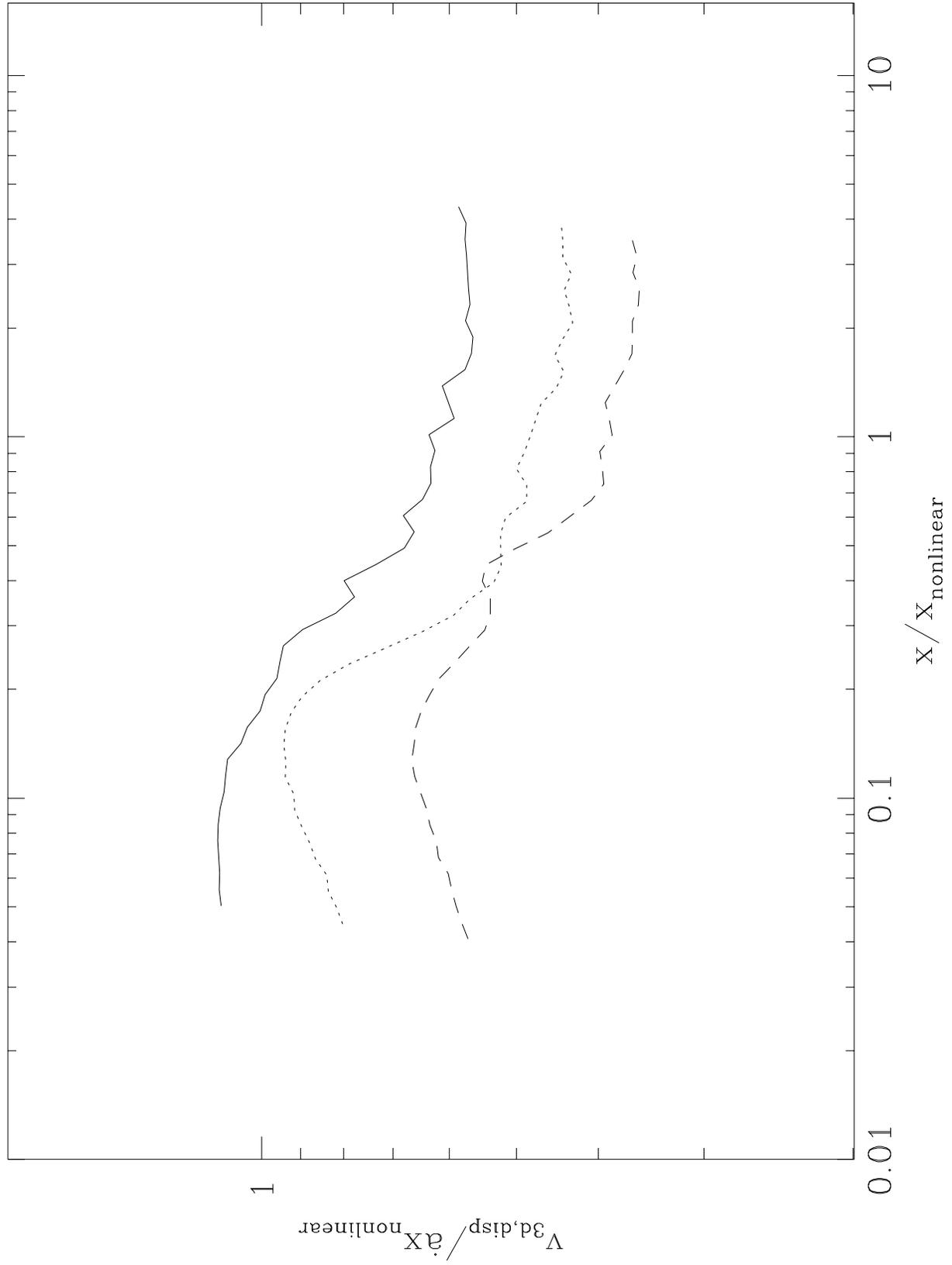
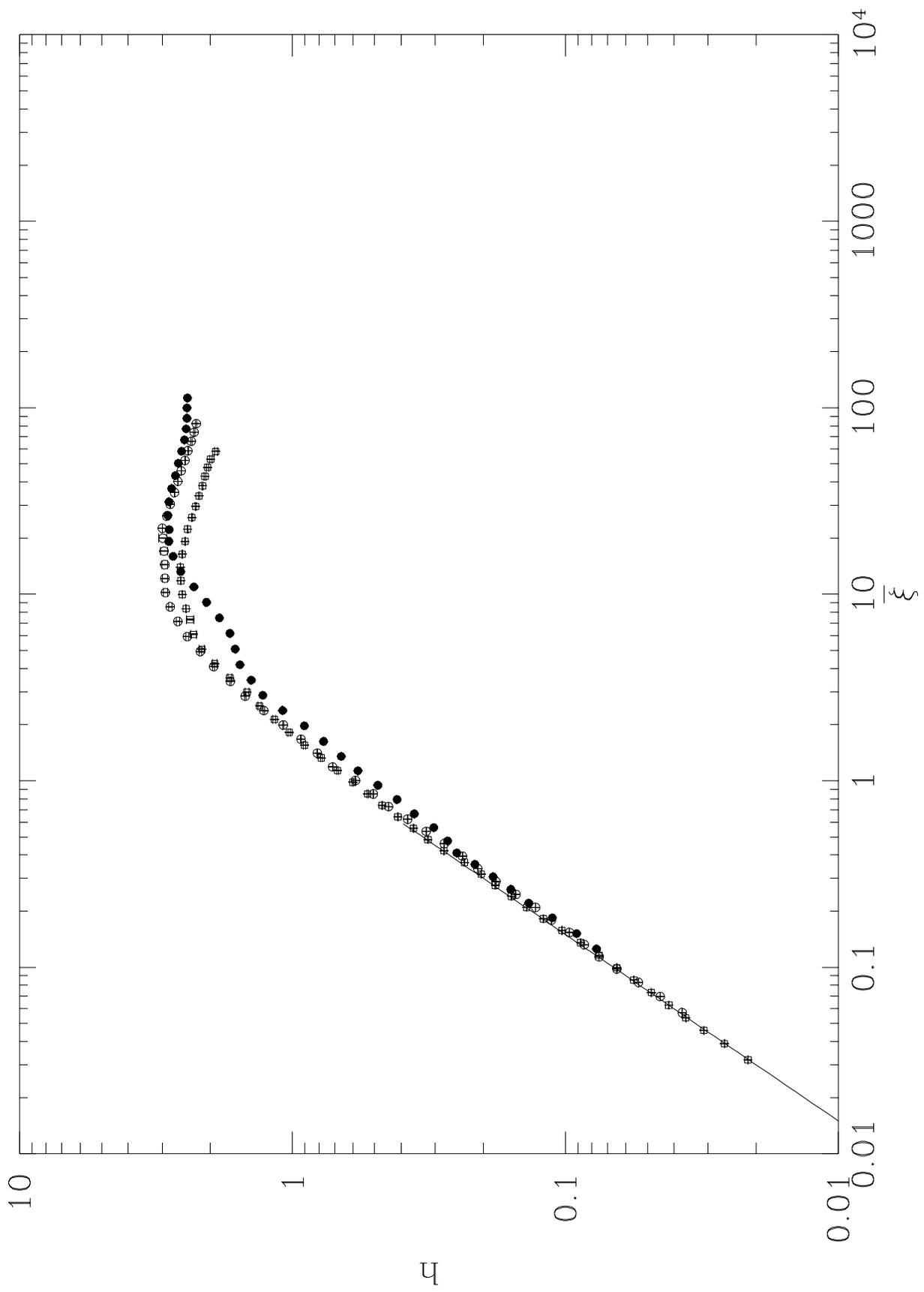
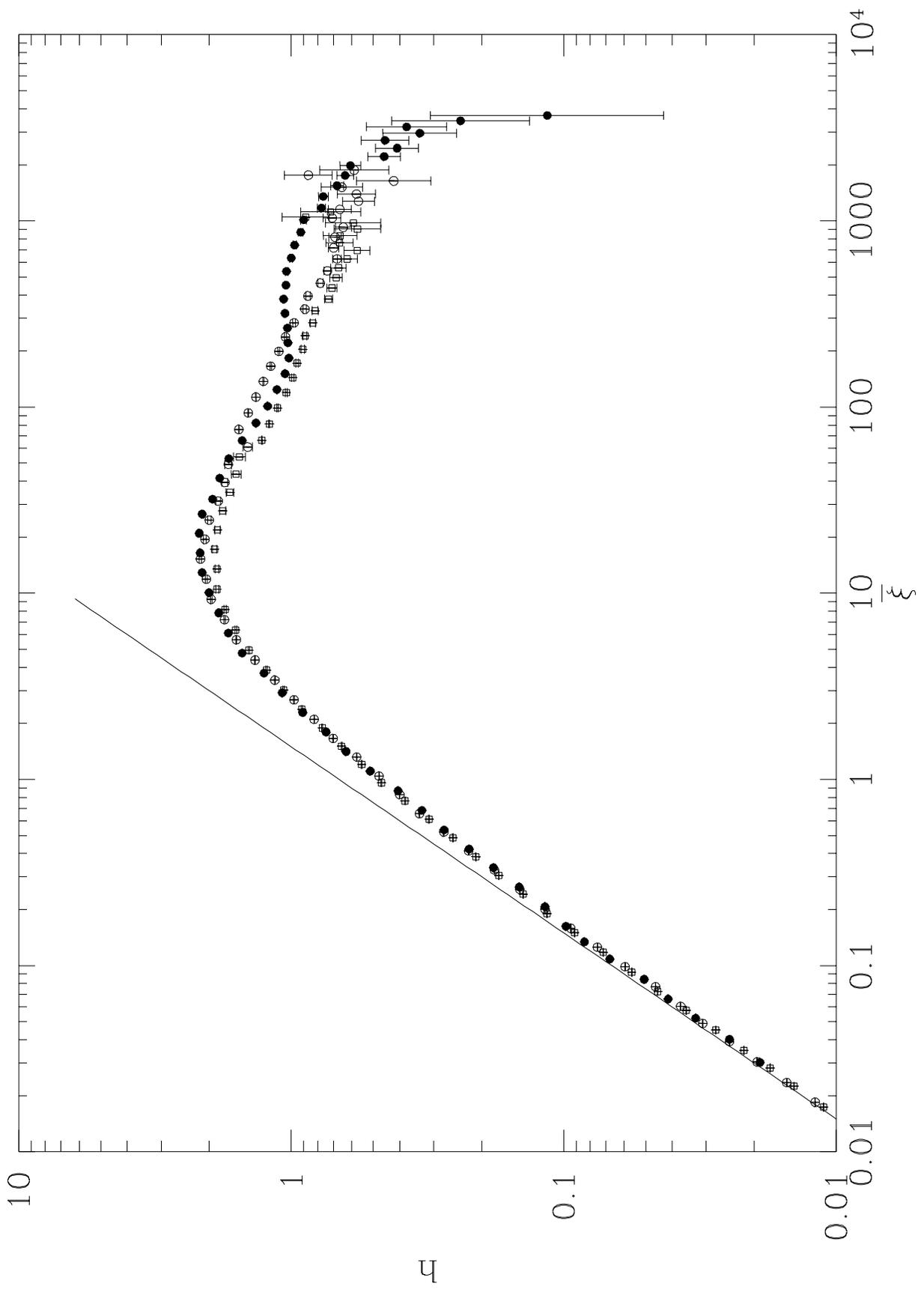
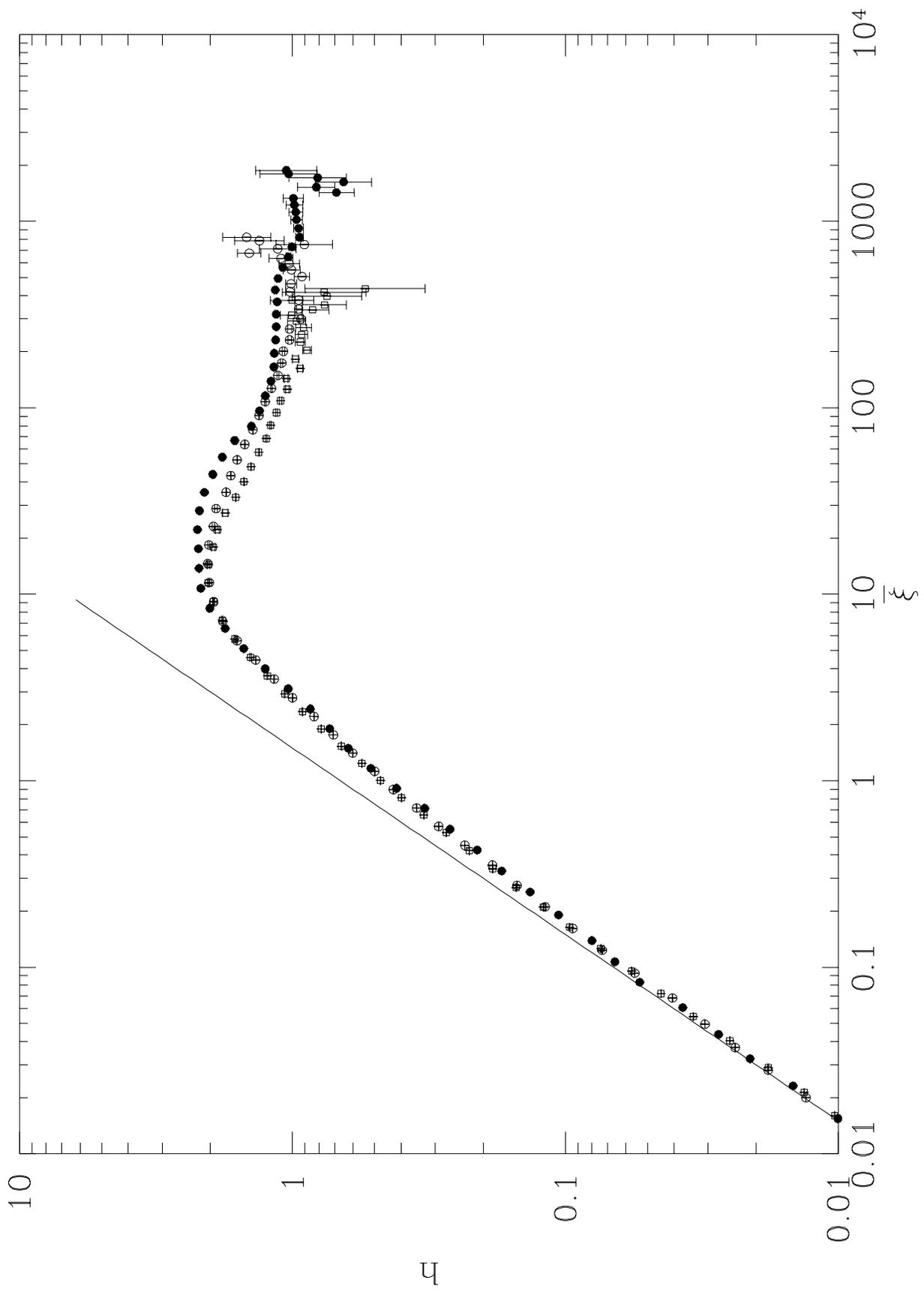
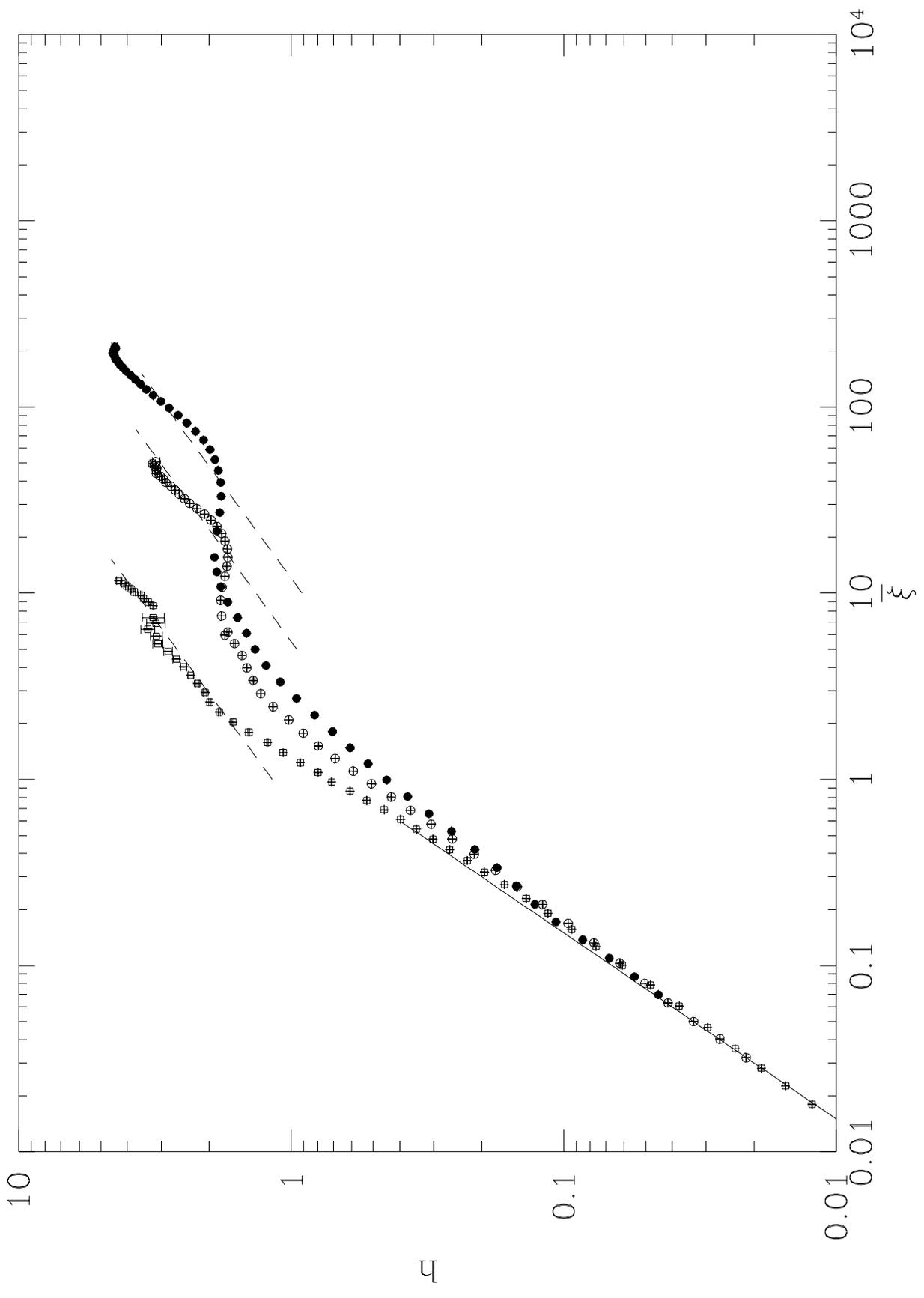
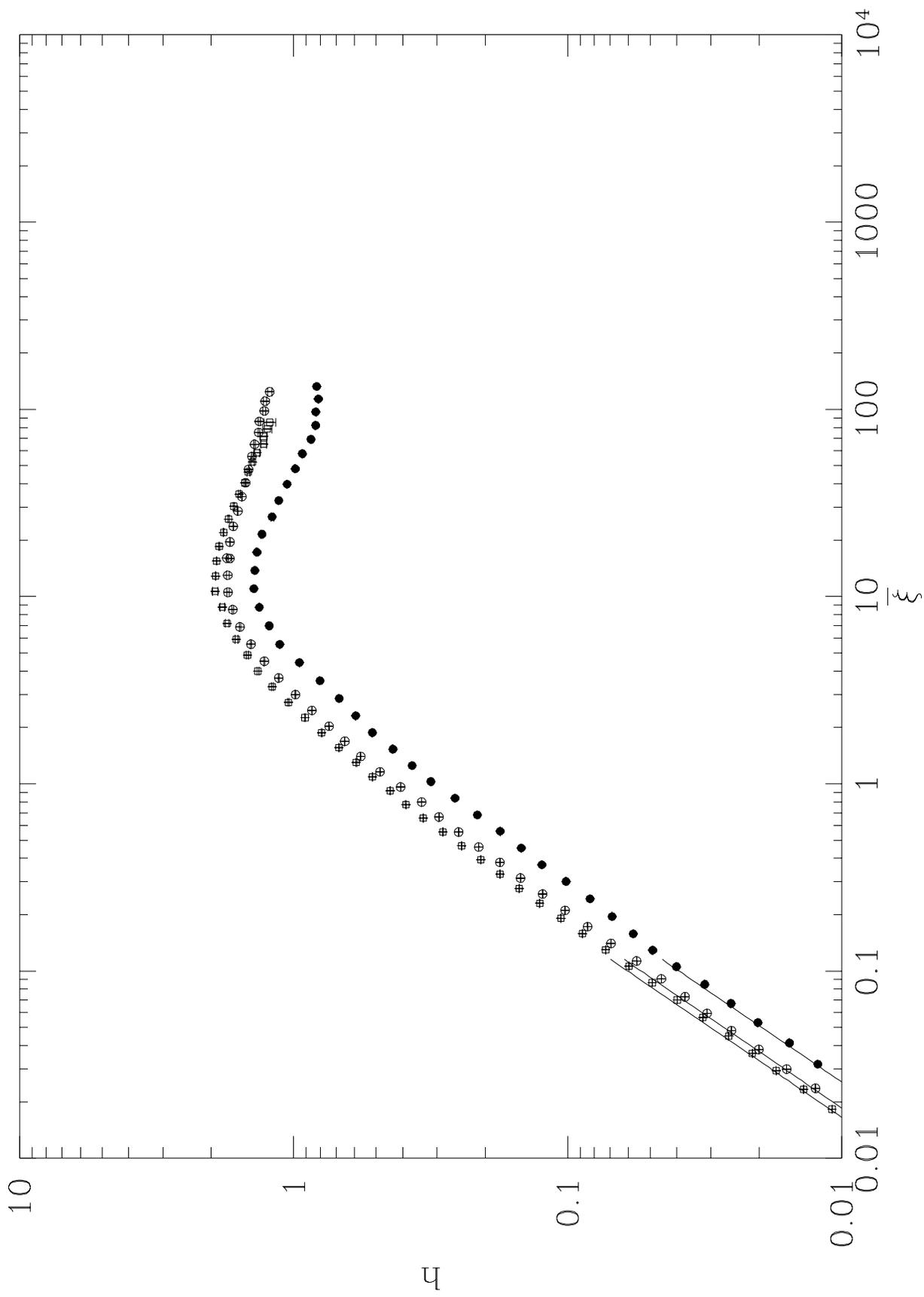
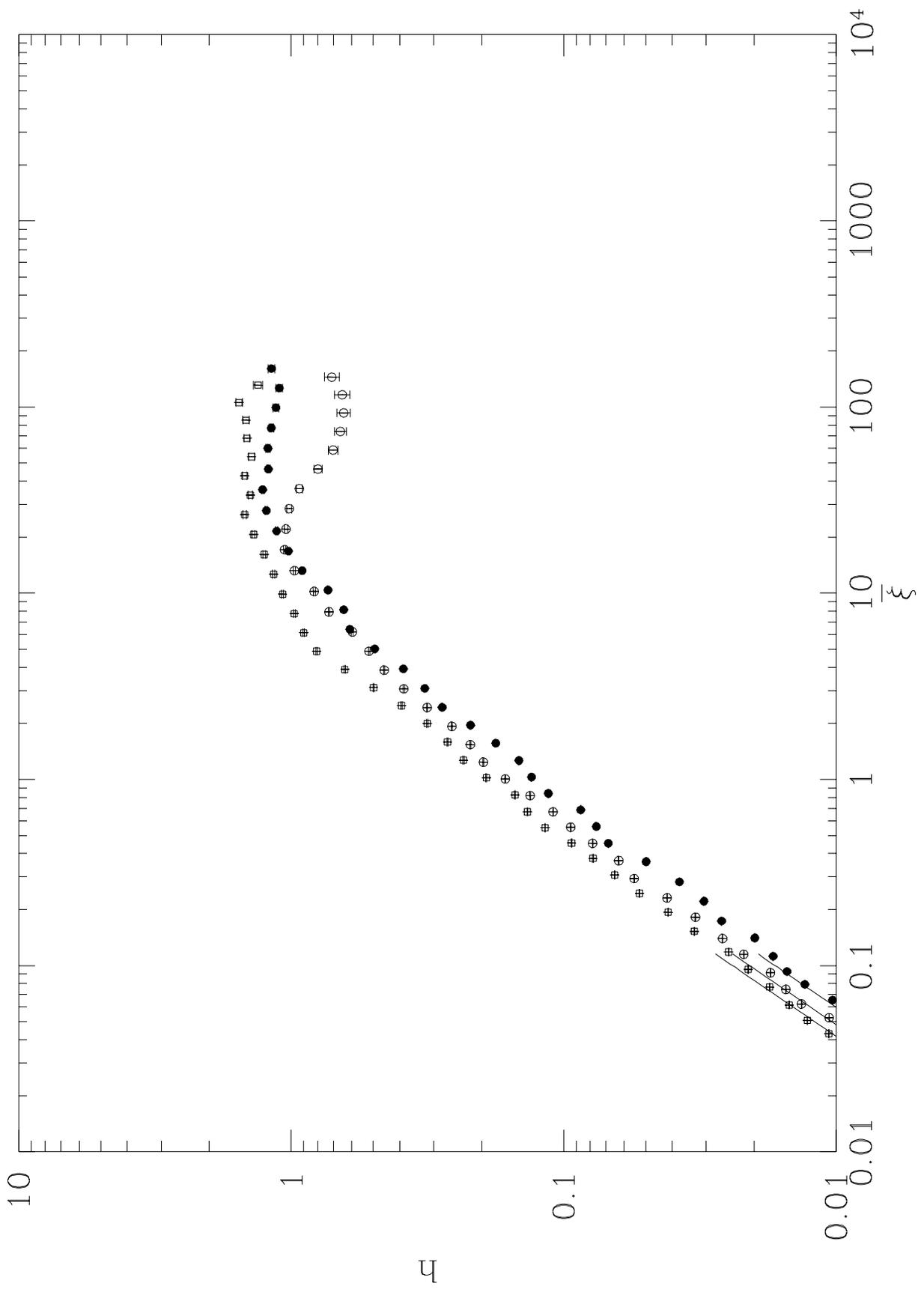
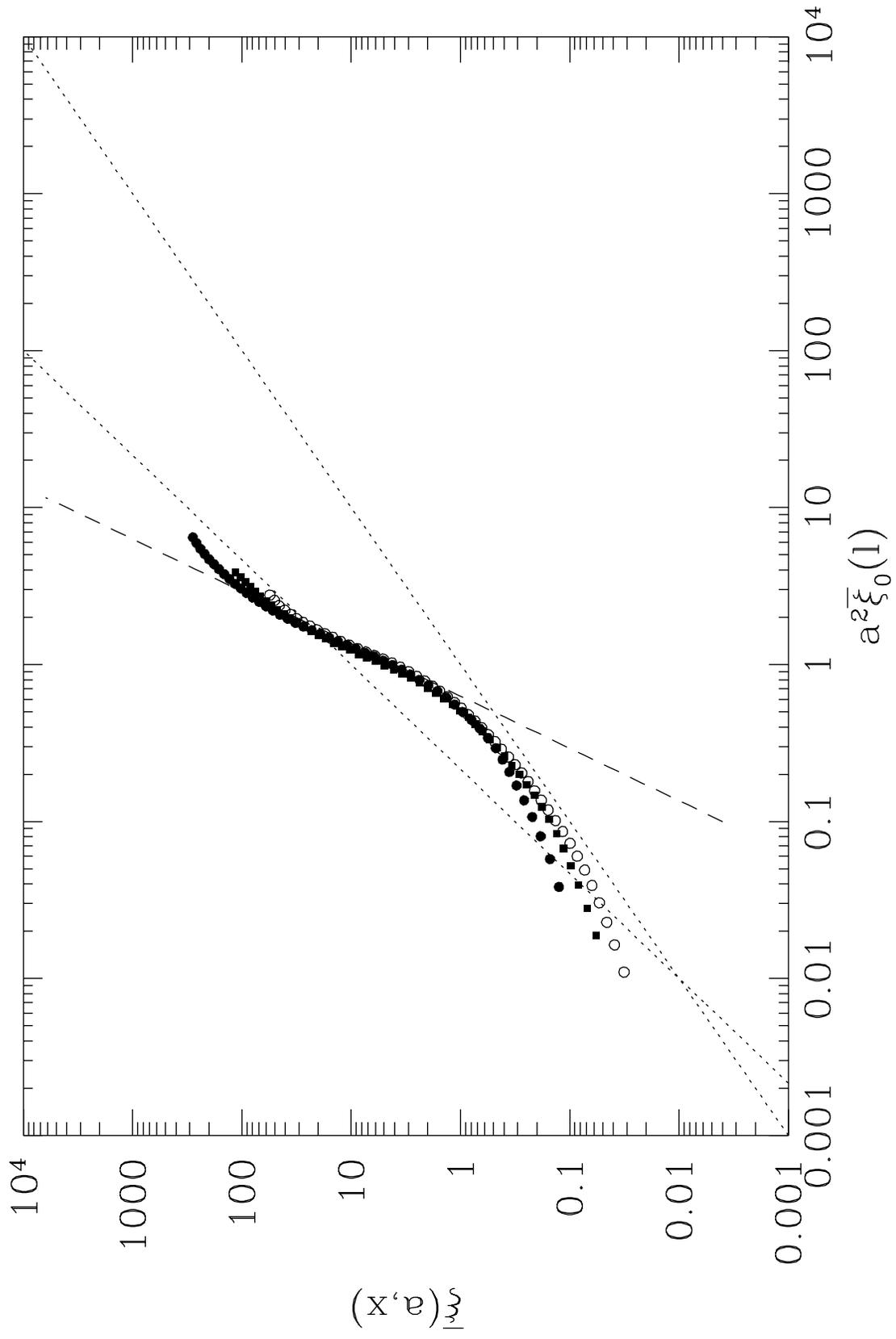
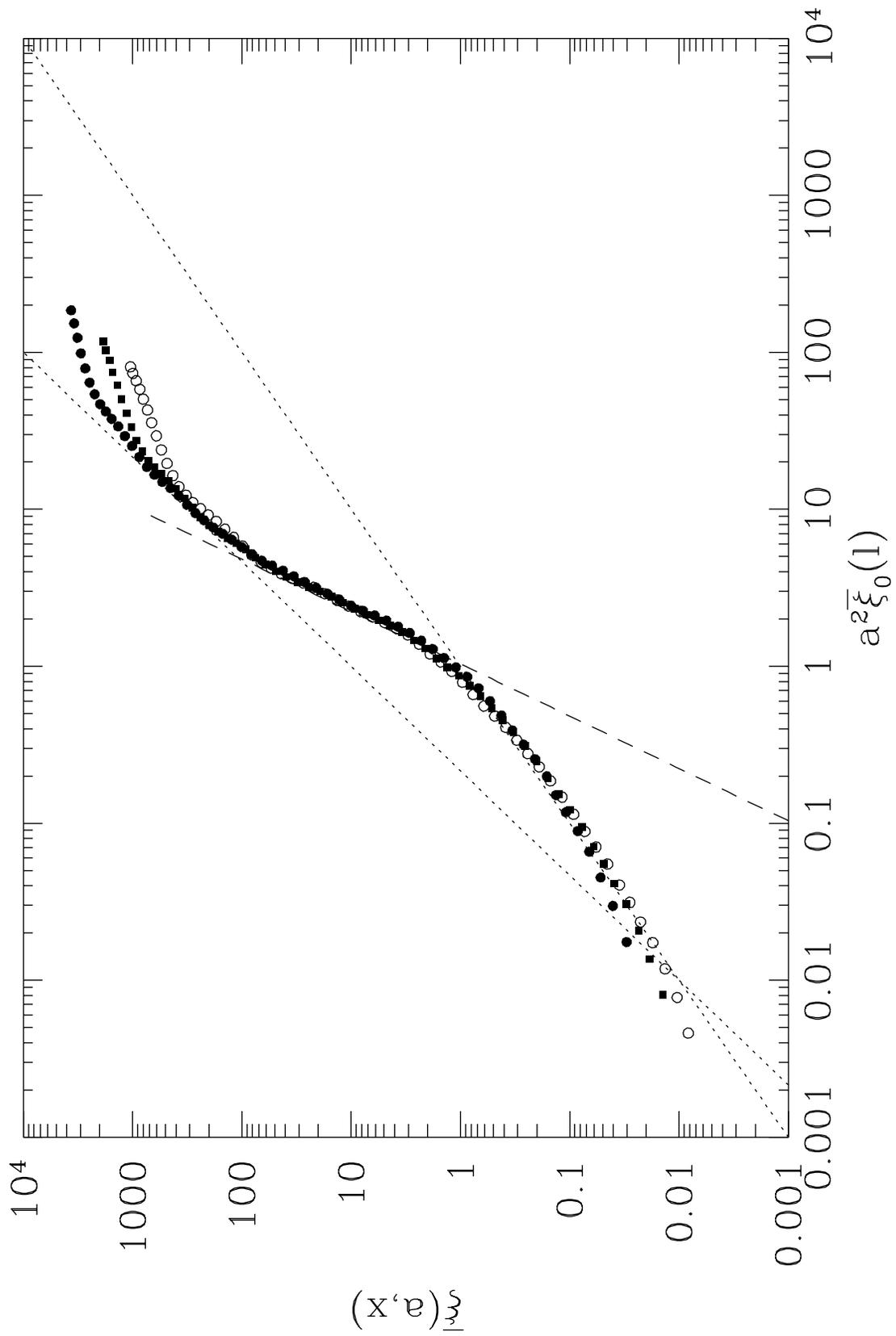
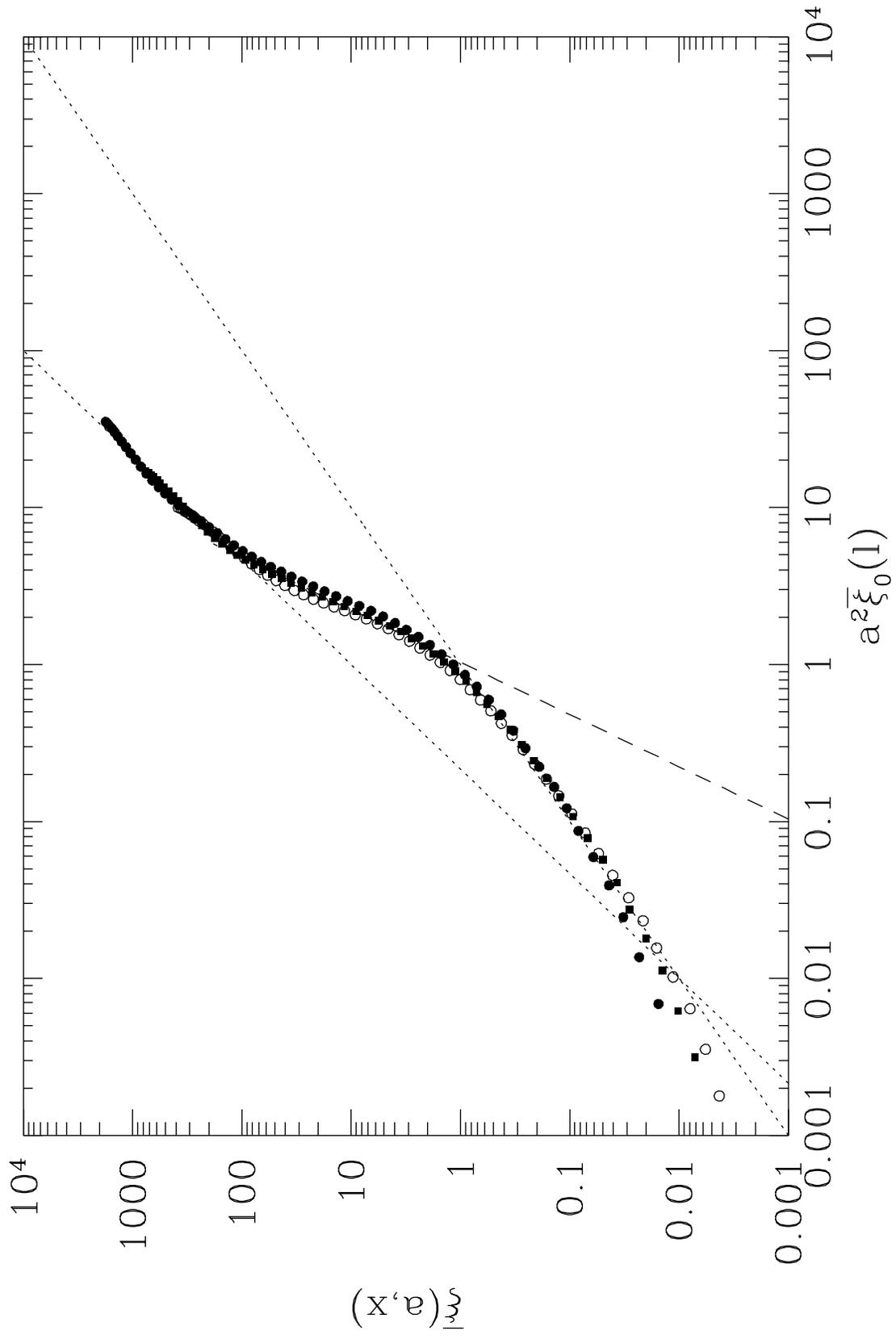
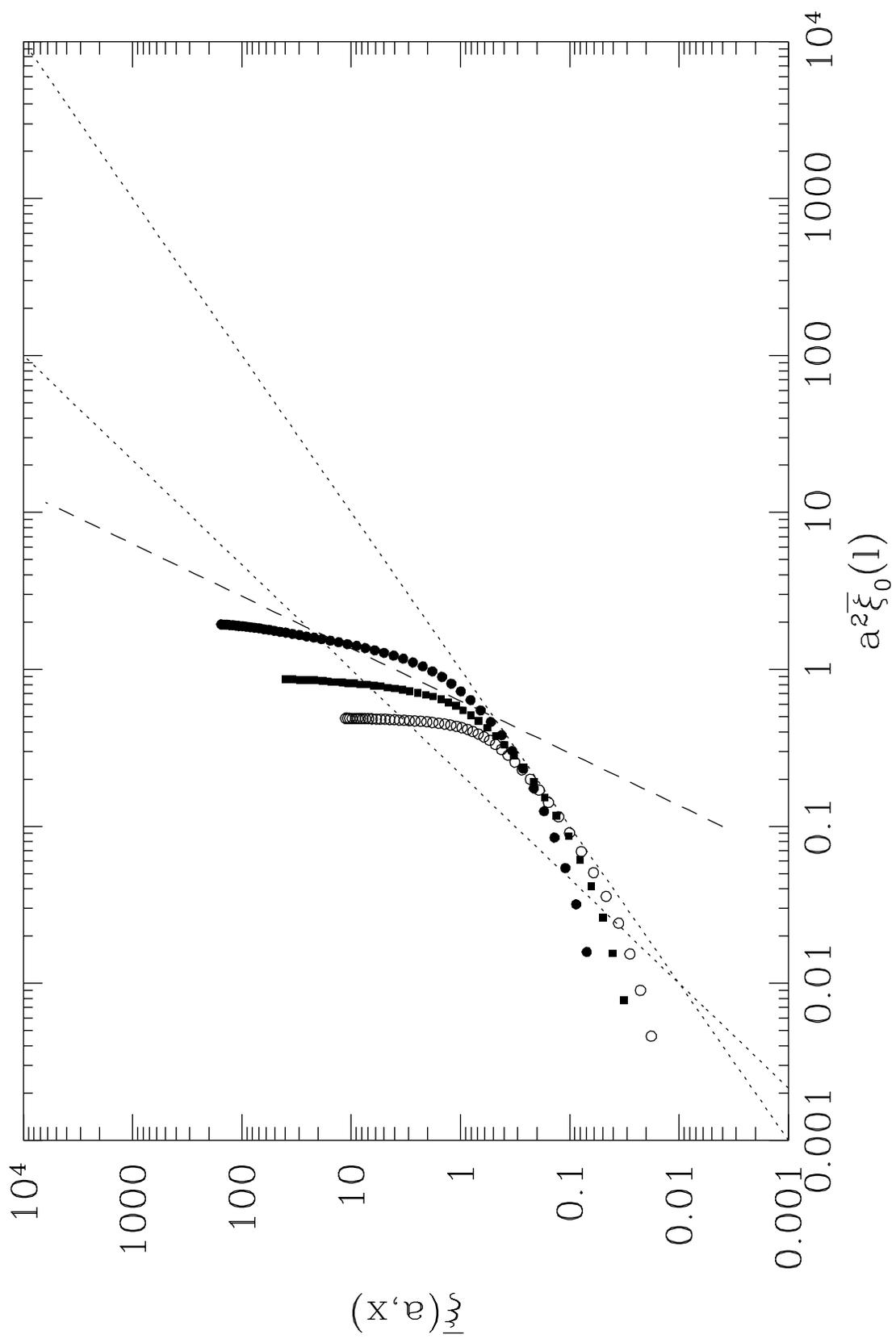
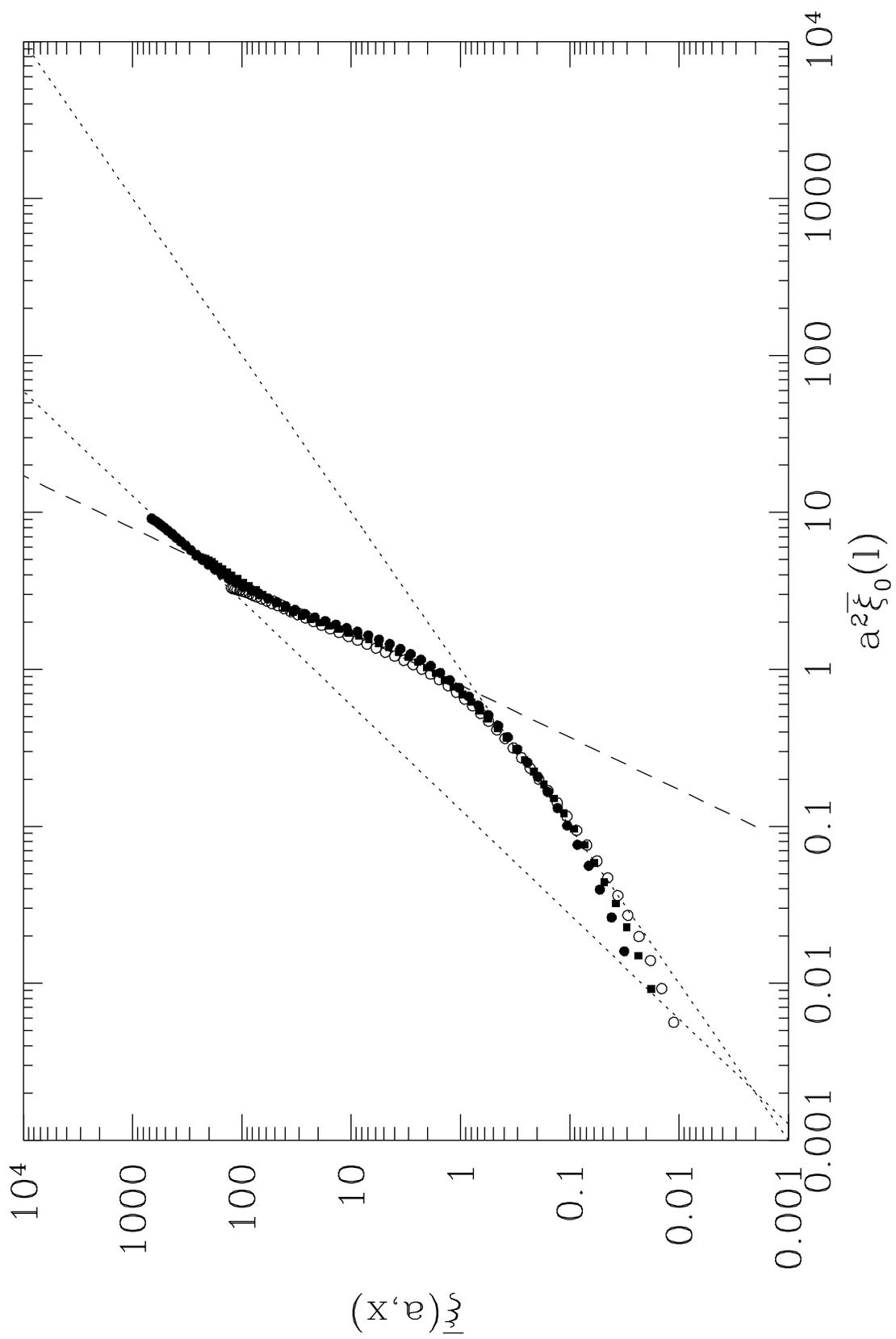
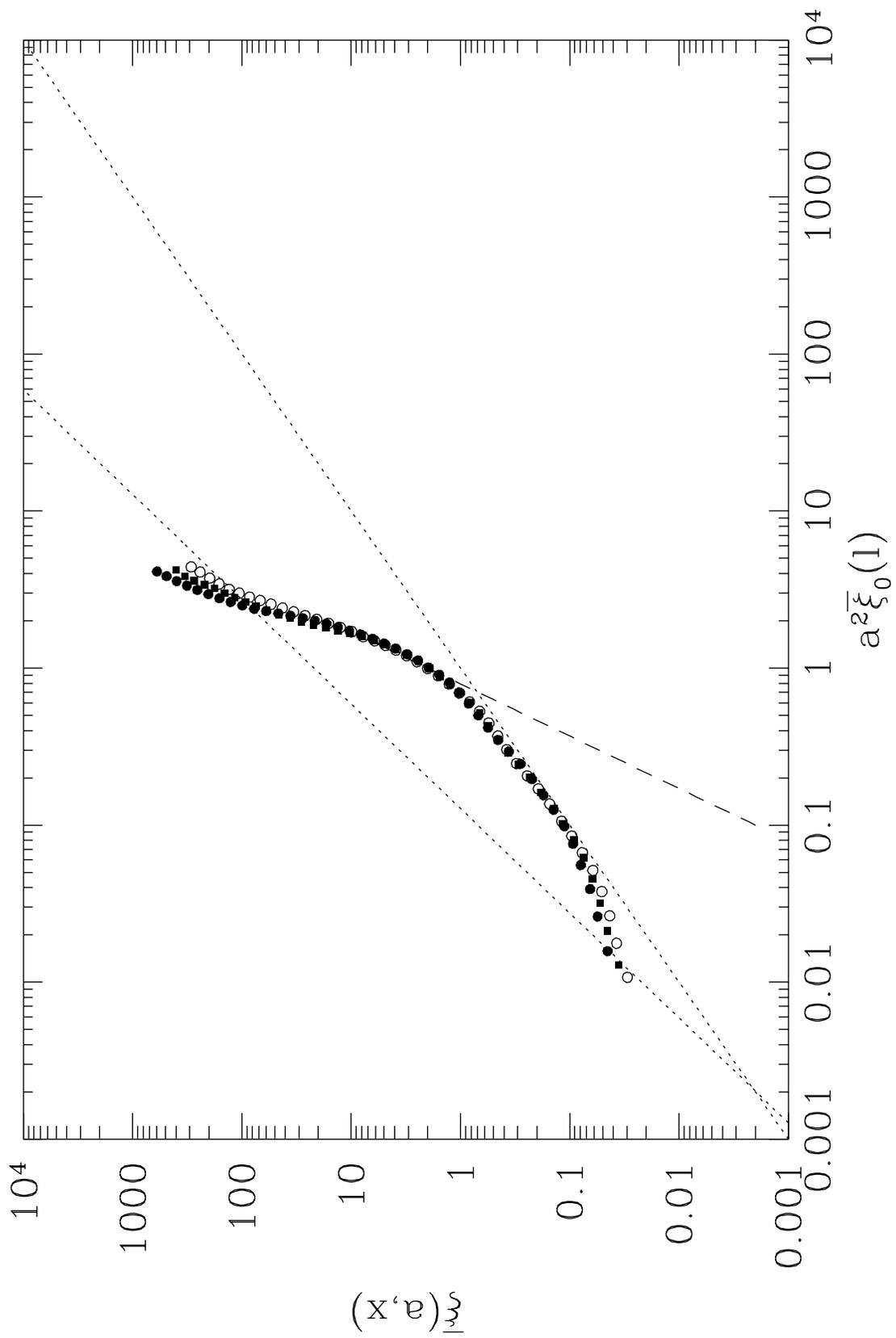
\end